\documentclass[11pt,draftcls,onecolumn]{IEEEtran} 
\usepackage{cite}
\usepackage{pslatex}
\usepackage{epsfig}
\usepackage{amsmath}
\usepackage{bm}
\usepackage{amssymb}
\usepackage{tabularx,framed,calc}
\usepackage{array}
\usepackage{threeparttable}
\usepackage{url}
\usepackage{dsfont}
\author{Chih Ning Hsu} 
\date{}  

\DeclareMathAlphabet{\mathpzc}{OT1}{pzc}{m}{it}

\begin{document}

\title{Joint Subcarrier Pairing and Power Allocation for OFDM Transmission with Decode-and-Forward Relaying}

\author{
\authorblockN{Chih-Ning Hsu, Hsuan-Jung Su* and Pin-Hsun Lin}\\
\authorblockA{Graduate Institute of Communication Engineering\\ Department of Electrical Engineering\\
National Taiwan University, Taipei, Taiwan\\
r96942060@ntu.edu.tw, hjsu@cc.ee.ntu.edu.tw, f89921145@ntu.edu.tw}}

\maketitle

\begin{abstract}
In this paper, a point-to-point Orthogonal Frequency Division Multiplexing (OFDM) system with a decode-and-forward (DF) relay is considered. The transmission consists of two hops. The source transmits in the first hop, and the relay transmits in the second hop. Each hop occupies one time slot. The relay is half-duplex, and capable of decoding the message on a particular subcarrier in one time slot, and re-encoding and forwarding it on a different subcarrier in the next time slot. Thus each message is transmitted on a pair of subcarriers in two hops. It is assumed that the destination is capable of combining the signals from the source and the relay pertaining to the same message. The goal is to maximize the weighted sum rate of the system by jointly optimizing subcarrier pairing
and power allocation on each subcarrier in each hop. The weighting of the rates is to take into account the fact that different subcarriers may carry signals for different services. Both total and individual power constraints for the source and the relay are investigated. For the situations where the relay does not transmit on some subcarriers because doing so does not improve the weighted sum rate, we further allow the source to transmit new messages on these idle subcarriers. To the best of our knowledge, such a joint optimization inclusive of the destination combining has not been discussed in the literature.
The problem is first formulated as a mixed integer programming problem. It is then transformed to a convex optimization problem by continuous relaxation, and solved
in the dual domain. Based on the optimization results, algorithms to achieve feasible solutions are also proposed.
Simulation results show that the proposed algorithms almost achieve the optimal weighted sum rate, and outperform the existing methods in various
channel conditions.
\end{abstract}

\vspace{-4mm}
\begin{center}
   {\underline{\bf \small EDICS}} \hspace{3mm} {\small WIN-RSMG}
\end{center}
\vspace{-6mm}

\section{Introduction}\label{introduction}
For an Orthogonal Frequency Division Multiplexing (OFDM) system with relay, identifying a proper way to
allocate resources to the source and the relay is the main
bottleneck for achieving good performance. In this paper, we consider a point-to-point OFDM system with a decode-and-forward (DF) half-duplex relay.
Each message is transmitted in two hops each occupying one time slot. A message transmitted by the source on one subcarrier in the first time slot is, if successfully decoded by the relay, forwarded by the relay to the destination on one (not necessarily the same) subcarrier in the second time slot. With the assumption that the channel state information (CSI) is known at
the source, many works have been done to make resource utilization of this system
more efficient.

A general downlink Orthogonal Frequency Division Multiple Access (OFDMA) relay system with
individual power constraints at one source and many relays was
considered in \cite{DF_individual_downlink}. In that work, joint
optimization of the subcarrier selection and power allocation was done. However, that work assumed that a message is received by a destination either directly from the source, or from a relay which forwarded the message. Destination combining of the signals directly from the source and forwarded by the relay pertaining to the same message was not considered. In addition, as each relay collectively uses its active subcarriers to forward messages to different destinations, a more complicated re-encoding scheme has to be used by the relay to fit the received message for a particular destination into the subcarriers designated to that destination.
In 
\cite{DF_OFDM_total_individual_power,relay_DF_individual_power_constraint,Vandendorpe_J}, optimal power
allocation for OFDM with DF relaying and fixed source and relay subcarrier pairing was proposed.
\cite{DF_OFDM_total_individual_power}\cite{Vandendorpe_J}
considered two kinds of power constraints: one is that the total transmit
power is shared between the source and the relay; the other has
individual power constraints for the source and the relay. In 
\cite{relay_eq_channel_gain,relay_AF_DF_eq_channel_gain,Li_AFDF_OFDM},
both power allocation and subcarrier pairing were considered for OFDM systems with relaying under the total power constraint. However, power allocation and subcarrier pairing were optimized separately.
\cite{relay_eq_channel_gain} proposed a subcarrier pairing method by sorting the subcarriers of the source-relay (SR) link and the relay-destination (RD) link, respectively, according to their channel gains. The SR subcarrier and the RD subcarrier with the same respective ranks are then paired together.
The optimality of this sorted channel pairing (SCP) scheme, in the absence of the source-destination (SD) link,
for both DF and amplify-and-forward (AF) relaying schemes were proved in \cite{relay_AF_DF_eq_channel_gain}\cite{Li_AFDF_OFDM}. SCP was also proposed in 
\cite{Herdin,wittneben,AF_pairing_without_diversity,relay_AF_OPT_SUBCHANNEL_ASSIGNMENT}
for OFDM AF relaying systems without the SD link, and in \cite{AF_OFDM_total_individual_power} when the SD link and destination combining are present. Power allocation with total and individual power constraints for OFDM AF relaying systems were considered in \cite{AF_pairing_without_diversity} and \cite{AF_OFDM_total_individual_power}, while \cite{wittneben} focused on only the total power constraint.
The above works dealing with power allocation for the OFDM AF relaying systems usually used approximations to relax the problem into a solvable one. Without making any approximations, \cite{relay_AF_total_optimal_power} investigated the optimal power allocation problem for the OFDM AF relaying systems with fixed subcarrier pairing and total power constraint in the absence of the SD link.

In view of the lack of joint optimization of power allocation and subcarrier pairing for OFDM systems with DF relaying in the literature, the goal of this paper is to solve this problem with the presence of the SD link and destination combining of signals from the source and the relay.
Both the total power constrained system and the individual power constrained system are considered. For the total power constrained system,
we formulate the joint power allocation and subcarrier pairing
problem as a mixed integer programming problem whose optimal solution is hard to
obtain. We then use some special properties of the system and the continuous relaxation \cite{DF_individual_downlink}\cite{concave_function} to reform the problem and solve the dual problem by the subgradient method \cite{subgradient_method}. With both the power and subcarrier pairing constraints, the optimization problem becomes very complicated, and the duality gap may not be zero. However, as verified by \cite{Yu_dual}\cite{zero_gap_SUB_infinity} and our own simulation, the duality gap is virtually zero when the number of subcarriers is reasonably large. Thus the dual optimum value becomes a very tight upper bound for the primal optimum for most practical systems. In addition to the duality gap, some other practical issues such as algorithm design and complexity comparison
are also discussed.
We then extend the
formulation to have individual power constraints, and find that the complications
caused by individual power constraints can be alleviated in the dual domain.
The dual optimum value is again a very tight upper bound for the primal optimum.

Finally, we relax the constraint that only the relay can transmit in the second time slot. Therefore, additional messages may be transmitted on the idle subcarriers in the SD link in the second time slot, when it is deemed that relaying on these subcarriers does not improve the weighted sum rate. Such a model was also considered in \cite{Vandendorpe_J}. However, \cite{Vandendorpe_J} optimized power allocation (and relaying modes) only for a particular subcarrier pairing scheme
without weighting of the rates. These conditions made the problem easier to solve.
In this paper, we consider joint optimization of power allocation and subcarrier pairing with weighted rates. The problem is more general and difficult. However, by defining an additional indicator, we can formulate the problem similarly as in the case without the second-slot SD transmission. The problem is then solved in the dual domain.
Simulation shows that, for this problem,
the duality gap is also nearly zero.

Based on the optimization results, algorithms to achieve feasible subcarrier pairing and power allocation are also proposed. Simulation results show that the proposed algorithms almost achieve the optimal weighted sum rate, and outperform the SCP proposed in \cite{relay_eq_channel_gain} in various
channel conditions.

The rest of this paper is organized as follows. Section \ref{Sec_system_model} describes the system model.
Section \ref{Sec_maximization_typeI} solves the optimization problem under the total power constraint.
Detailed discussions on the practical issues are also presented in this section.
Section \ref{sec_maximization_individual} solves the optimization problem under the individual power constraints.
Section \ref{sec_maximization_total_extra_direct} formulates and solves the optimization problem for the system with additional messages transmitted on the SD link in the second time slot, under both total and individual power constraints. Section \ref{Sec_simulation} summarizes our results and observations. Section \ref{Sec_conclusion} concludes this paper.

\section{System Model}\label{Sec_system_model}
We consider a two-hop DF relay system consisting of one source,
one relay, and one destination. OFDM with the same spectral occupancy is used for all links. The
total frequency band is divided into $M$ subcarriers. To avoid interference, for
each subcarrier, only one node (the source or
the relay) transmits in a given time slot. All time slots are of the same duration. The source transmits in the first time slot while the relay and the destination receive. The relay is half-duplex that receives in the first time slot and transmits in the second time slot. Each subcarrier used by the source in the first time slot is paired with one subcarrier used by the relay in the second
time slot to convey a message. Therefore the number of subcarrier pairs in transmission is $M$. If subcarrier $k$ in the first time slot and subcarrier
$m$ in the second time slot are paired, we call them subcarrier pair (SP) $(k,m)$. It is assumed that the relay re-encodes the received message with the same codebook as the one used by the source. The destination maximum ratio combines (MRC) the signals from the source in the first time slot and from the relay in the second time slot pertaining to the same message to exploit the spatial diversity.
The messages transmitted on different SPs are assumed to be independent.

\begin{figure}[htp]
\centering
\includegraphics[width=0.55\textwidth]{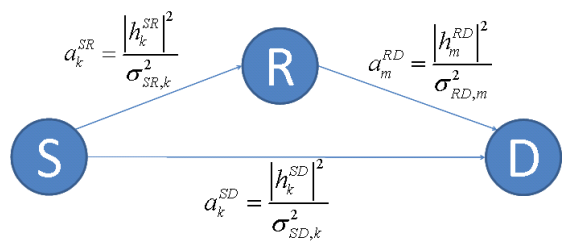}
\caption{Channel model for subcarrier pair $(k,m)$.}
\label{Fig_channel_model}
\end{figure}

The channel model associated with SP $(k,m)$
is shown in Fig. \ref{Fig_channel_model}. We use $h^{SD}_{k}$,
$h^{SR}_{k}$, and $h^{RD}_{m}$ to denote the channel gains of the
SD link, SR link,
and RD link
on subcarriers $k$, $k$, and $m$, respectively. $\sigma^2_{SD,k}$, $\sigma^2_{SR,k}$,
and $\sigma^2_{RD,m}$ are the variances of the additive white Gaussian
noises (AWGN) in the corresponding channels. As shown in Fig.
\ref{Fig_channel_model}, we use $a^{SD}_{k}=\frac{| h^{SD}_{k}|^2}{\sigma^2_{SD,k}}$,
$a^{SR}_{k}=\frac{|h^{SR}_{k}|^2}{\sigma^2_{SR,k}}$, and $a^{RD}_{m}=\frac{|h^{RD}_{m}|^2}{\sigma^2_{RD,m}}$ to denote the normalized channel gains.
The channels are assumed to remain constant in a two-slot period. All the normalized channel gains are assumed known at the source which will perform subcarrier pairing and power allocation. The source then informs the relay and the destination of the corresponding parameters via proper control signaling before the data transmission. These assumptions are reasonable for the situations where the channel coherence time is longer than the sum of the CSI measurement and feedback time, the control signaling time, and the data transmission duration.

In practical implementation, the channel gains can be measured at the relay and the destination during the training period preceding the data transmission period. The training period has a similar structure as the data transmission period in which the source transmits training signals during the first time slot while the relay and the destination measure the SR and SD channels, respectively. The relay then transmits training signals in the second time slot to let the destination measure the RD channel. A training slot could be shorter than a data transmission slot. The measured channel gains can be fed back to the source on dedicated reverse control channels. After the source has done subcarrier pairing and power allocation, it can embed the pairing and power allocation parameters in the beginning of the first-slot data transmission. This embedded control signal is transmitted with stronger power and/or more reliable coding. So it can be guaranteed that the relay and destination can successfully decode the relevant parameters to figure out how to receive (and for the relay, how to forward as well) the upcoming data.

Taking the same assumption as in 
\cite{DF_OFDM_total_individual_power,relay_DF_individual_power_constraint,relay_eq_channel_gain,relay_AF_DF_eq_channel_gain,Li_AFDF_OFDM},
in Section~\ref{Sec_maximization_typeI} and Section~\ref{sec_maximization_individual} we first consider the scenario where for each SP $(k,m)$, the source only transmits in the first time slot. Even if it is decided that the relay will not transmit on subcarrier $m$, the source is not allowed to use this idle subcarrier in the second time slot. In Section~\ref{sec_maximization_total_extra_direct}, this restriction is relaxed and the source is allowed to transmit additional messages in the second time slot on the subcarriers not used by the relay. This model has also been investigated in \cite{Vandendorpe_J} which assumed fixed subcarrier pairing with SPs $(k,k), k=1, 2, \ldots, M$. Together with unweighted rates, the $(k,k)$ subcarrier pairing makes determination of whether the relay will be active for SP $(k,k)$ and optimal power allocation among the SPs easier to solve. However, it is inferior and less general than the joint optimization of subcarrier pairing and power allocation considered in Section~\ref{sec_maximization_total_extra_direct}.

For the sake of generality, we consider weighted sum rate as the performance metric. A weighting factor $w_{k} \geq 0$ is assigned to
the rate transmitted by the source on subcarrier $k$ to reflect different priorities or quality-of-service (QoS) requirements.

\section{Weighted Sum Rate Maximization under Total Power Constraint}\label{Sec_maximization_typeI}
In this section, we consider joint optimization of subcarrier pairing and power allocation to achieve the highest weighted sum rate under the total power constraint. We first give the problem formulation. Then a solution in the dual domain is given. The duality gap and achievability of the optimal solution, together with some practical algorithm design issues, will be discussed.

\subsection{Primal Problem Formulation}
\label{primal_total}
For a given SP $(k,m)$, let $R_{k,m}$ be its achievable weighted rate, and $p^{S}_{k,m}$ and $p^{R}_{k,m}$ be the source power
in the first time slot and the relay power in the second time slot, respectively.
Depending on whether the relay is active, this SP may work in either the relay mode or the direct-link mode.
In the relay mode, the half-duplex relay forwards the message on subcarrier $m$ in the second time slot. In the direct-link
mode, the relay does not forward, and only subcarrier $k$ of the SD link
in the first time slot is used to transmit the message. Thus the  weighted rate achievable with Gaussian codebooks for SP $(k,m)$ can be expressed as \cite{relay_capacity_theorem}
\begin{equation}\label{Eq_pair_rate_typeI}
R_{k,m}=\left\{
\begin{aligned}
&\frac{w_{k}}{2}\log(1+a^{SD}_{k}p^{S}_{k,m}), &\text{direct-link mode}\\
&\frac{w_{k}}{2}\min\left\{\log\left(1+a^{SR}_{k}p^{S}_{k,m}\right),\; \log\left(1+a^{SD}_{k}p^{S}_{k,m}+a^{RD}_{m}p^{R}_{k,m}\right)\right\}, &\text{relay mode}\\
\end{aligned}
\right.
\end{equation}
where the rate is
scaled by $\frac{1}{2}$ because the transmission takes two time slots.

Under the total power constraint of $p_{k,m}=p_{k,m}^S+p_{k,m}^R$ for the SP $(k,m)$, using relay is advantageous in terms of maximizing the achievable rate when \cite{DF_OFDM_total_individual_power}
\begin{equation}\label{relay_condition_total}
a^{SR}_{k}>a^{SD}_{k}~ \text{ and } ~a^{RD}_{m}>a^{SD}_{k}.
\end{equation}
In addition, based on the fact that, for the relay mode, the achievable rate is
maximized when the amounts of received information at the relay and the destination are the same, the expressions in
(\ref{Eq_pair_rate_typeI}) can be unified as \cite{relay_eq_channel_gain}
\begin{equation}\label{unified_rate}
\begin{aligned}
R_{k,m}=\frac{w_k}{2}\log(1+a_{k,m}p_{k,m}).
\end{aligned}
\end{equation}
This is obtained by letting
\begin{equation}\label{power_ratio}
\begin{aligned}
&p^{S}_{k,m}=\left\{
\begin{aligned}
&\frac{a^{RD}_{m}}{a^{SR}_{k}+a^{RD}_{m}-a^{SD}_{k}}p_{k,m},&\text{relay mode}\\
&p_{k,m},&\text{direct-link mode}\\
\end{aligned}
\right.\\
&p^{R}_{k,m}=\left\{
\begin{aligned}
&\frac{a^{SR}_{k}-a^{SD}_{k}}{a^{SR}_{k}+a^{RD}_{m}-a^{SD}_{k}}p_{k,m},&\text{relay mode}\\
&0,&\text{direct-link mode}
\end{aligned}
\right.
\end{aligned}
\end{equation}
in (\ref{Eq_pair_rate_typeI}), and defining $a_{k,m}$ as the equivalent channel gain given by
\begin{equation}\label{Eq_ECG}
a_{k,m}=\left\{
\begin{aligned}
&\frac{a^{SR}_{k}a^{RD}_{m}}{a^{SR}_{k}+a^{RD}_{m}-a^{SD}_{k}},&\text{relay mode}\\
&a^{SD}_{k},&\text{direct-link mode}.
\end{aligned}
\right.
\end{equation}
Thus, when the channel gains are known, for any possible pairing, whether a SP $(k,m)$ should be in the relay mode or the direct-link mode, and the maximum achievable weighted rate of this SP as a function of the total power $p_{k,m}$, can be derived immediately.
Define an indicator $t_{k,m}$ which
is 1 if SP $(k,m)$ is selected, and 0 otherwise.
The weighted sum rate optimization problem can be formulated as
\begin{align}
\label{objective_Rate_I_MIP}
\max_{{\bm p},{\bm t}}\text{ }&\sum^M_{k=1}\sum^M_{m=1}t_{k,m}\frac{w_{k}}{2}\log(1+a_{k,m}p_{k,m})
\\
\label{constraint_power_consumption}
\text{s.t.}\text{ }&\sum^M_{k=1}\sum^M_{m=1}p_{k,m}\leq P,
\\
\label{constraint_t_k}
&\sum^M_{k=1}t_{k,m}=1,\text{ }\forall m,
\\
\label{constraint_t_m}
&\sum^M_{m=1}t_{k,m}=1,\text{ }\forall k,
\\
\label{constraint_power_0}
&p_{k,m}\geq~0,\text{ }\forall k, m,
\\
\label{constraint_t_binary}
&t_{k,m}\in\{0,1\},\text{ }\forall k, m,
\end{align}
where $P$ is the total power constraint,
${\bm p}\in \mathds{R}_+^{M \times M}$ (with $\mathds{R}_+$ denoting the set of nonnegative real numbers) and
${\bm t}\in\{0,1\}^{M \times M}$ are matrices with entries $p_{k,m}$
and $t_{k,m}$, respectively. Since the power allocated to the unselected SPs does not contribute
to the weighted sum rate, it is obvious that the optimal solution will only allocate non-zero
power to the selected SPs. Although similar in the approach, there are some significant differences between the above problem formulation and the ones in
\cite{DF_individual_downlink} and \cite{relay_AF_OPT_SUBCHANNEL_ASSIGNMENT}.
\cite{DF_individual_downlink} and \cite{relay_AF_OPT_SUBCHANNEL_ASSIGNMENT}
both did not consider the SD link and destination combining when the relay is used. In
\cite{relay_AF_OPT_SUBCHANNEL_ASSIGNMENT}, the power allocated to
each subcarrier is fixed. As mentioned in Section \ref{introduction},
the relays in \cite{DF_individual_downlink} have to use complicated re-encoders with codebooks different from that of the source.
These differences make our optimization problem distinct from \cite{DF_individual_downlink} and \cite{relay_AF_OPT_SUBCHANNEL_ASSIGNMENT}.

The above problem is a mixed integer programming (MIP) problem
which is hard to solve. Therefore, as in \cite{concave_function}\cite{multilevel_waterfilling}, we relax the integer constraint
of (\ref{constraint_t_binary}) as $t_{k,m}\in \mathds{R}_+, \forall k, m$. This continuous relaxation makes $t_{k,m}$ the time sharing
factor of each SP.
The relaxed problem then becomes
\begin{align}
\label{objective_Rate_I_relaxed}
\max_{{\bm p},{\bm t}} ~\frac{1}{2}\sum^M_{k=1}\sum^M_{m=1}t_{k,m}~w_{k}\log\left(1+a_{k,m}\frac{p_{k,m}}{t_{k,m}}\right)~~\text{s.t.}~~
(\ref{constraint_power_consumption}),
(\ref{constraint_t_k}), (\ref{constraint_t_m}),
(\ref{constraint_power_0}), ~\text{and}\\
\label{constraint_t_0}
t_{k,m}\geq 0,\text{ }\forall k, m.
\end{align}
Note that the value of the objective function (\ref{objective_Rate_I_relaxed}) is the same as that of the original
objective function (\ref{objective_Rate_I_MIP}) when $t_{k,m}\in\{0,1\}, \forall k, m$.
This objective function is concave
because it is a nonnegative weighted sum of concave functions in the form of $x\log(1+\frac{y}{x})$ which is concave in $(x, y)$
\cite{concave_function}.
Since (\ref{objective_Rate_I_relaxed}) is a standard
convex programming problem, it can be solved by numerical
search algorithms such as the interior-point method
\cite{Book_Boyd}. However, the optimal $t_{k,m}$ may not be
integer-valued. Therefore, we opt to solve this problem by the dual
method which can provide an upper bound for problem
(\ref{objective_Rate_I_relaxed}) (by the weak duality \cite{Book_Boyd}).
In Section \ref{sec_dual_type_I_total}, it will be shown that the solution obtained by the dual method has $t_{k,m}\in\{0,1\}, \forall k, m$.

\subsection{Dual Problem} \label{sec_dual_type_I_total}
By dualizing constraints (\ref{constraint_power_consumption}) and
(\ref{constraint_t_k}), we obtain the Lagrangian as
follows:
\begin{equation}\label{Lagrangian_typeI}
\begin{aligned}
L( {\bm p}, {\bm t},\mu, {\bm \alpha})=&\frac{1}{2}\sum_{k=1}^M\sum_{m=1}^Mt_{k,m}~w_{k}\log\left(1+a_{k,m}\frac{p_{k,m}}{t_{k,m}}\right)\\
&+\mu\left(P-\sum_{k=1}^M\sum_{m=1}^M~p_{k,m}\right)+\sum_{m=1}^M\alpha_m\left(1-\sum_{k=1}^M~t_{k,m}\right),
\end{aligned}
\end{equation}
where $\mu\in \mathds{R}_+$ and ${\bm \alpha} ~(\text{the vector of} ~\alpha_m) \in\mathds{R}^M$ are the dual variables, with $\mathds{R}$ denoting the set of real numbers.
The dual objective function is
\begin{equation} \label{P_Lagrange_typeI}
h(\mu, {\bm \alpha})=
\max_{{\bm p}, {\bm t}} ~L({\bm p},{\bm t},\mu, {\bm \alpha}) ~~\text{s.t.}~~ (\ref{constraint_t_m}),(\ref{constraint_power_0}),(\ref{constraint_t_0})
\end{equation}
and the dual problem is
\begin{equation} \label{objective_dual}
\min_{\mu, {\bm \alpha}} ~h(\mu, {\bm \alpha})
~~\text{s.t.}~~ \mu\geq 0.
\end{equation}
It is well known that a function can be maximized by
first maximizing over some of the variables, and then maximizing
over the remaining ones \cite[Sec 4.1.3]{Book_Boyd}.
Thus we first solve $p_{k,m}$ for (\ref{P_Lagrange_typeI}) by
\begin{equation}\label{Eq_L_partial_p_typeI}
\frac{\partial{L}}{\partial{p_{k,m}}}=\frac{t_{k,m}w_{k}}{2}\frac{\frac{a_{k,m}}{t_{k,m}}}{1+a_{k,m}\frac{p_{k,m}}{t_{k,m}}}-\mu=\frac{w_{k}}{2}\frac{1}{\frac{1}{a_{k,m}}+\frac{p_{k,m}}{t_{k,m}}}-\mu =0
\end{equation}
with constraint (\ref{constraint_power_0}). The optimal solution is
\begin{align}\label{optimal power_typeI}
p_{k,m}^\ast=t_{k,m}\left[\frac{w_{k}}{2\mu}-\frac{1}{a_{k,m}}\right]^+,
\end{align}
where $x^+\triangleq\max\{x,0\}$. This is similar to the result of
multi-level water-filling\cite{multilevel_waterfilling}. We then
rewrite (\ref{Lagrangian_typeI}) as
\begin{equation}
\begin{aligned}
L({\bm p}^*,{\bm t},\mu, {\bm \alpha})=\sum_{k=1}^M\sum_{m=1}^M~t_{k,m}X_{k,m}+K(\mu, {\bm \alpha}),
\end{aligned}
\end{equation}
where
\begin{equation}\label{compute X}
X_{k,m}=\frac{w_{k}}{2}\log\left(1+a_{k,m}\left[\frac{w_{k}}{2\mu}-\frac{1}{a_{k,m}}\right]^+\right)-\alpha_m
-\mu\left(\left[\frac{w_{k}}{2\mu}-\frac{1}{a_{k,m}}\right]^+\right),
\end{equation}
\begin{equation}
K(\mu, {\bm \alpha})=\mu P+\sum_{m=1}^M \alpha_m.
\end{equation}
We give an intuitive explanation for each term in $X_{k,m}$. The
first term can be viewed as the rate obtained by selecting
subcarrier $m$ in the second time slot for subcarrier
$k$ in the first time slot. $\alpha_m$ is the penalty
of selecting subcarrier $m$ in the second time slot. The
last term is the price of power consumption.

Due to the fact that $K(\mu, {\bm \alpha})$ and $X_{k,m}$ are independent
of ${\bm t}$, we can easily find the optimal
${\bm t}$ for (\ref{P_Lagrange_typeI}) with constraints (\ref{constraint_t_m}) and (\ref{constraint_t_0}) as
\begin{equation}\label{optimal_t_typeI}
t_{k,m}^\ast=\left\{
\begin{aligned}
&1,\text{ }m=\arg\max_{m=1,...,M}X_{k,m}\\
&0,\text{ otherwise}
\end{aligned}, ~~~~\forall k
\right. .
\end{equation}
In operation, we first assume that $\mu$ and $\alpha_m$'s are given. Then the power allocation for every possible SP can be computed by (\ref{optimal power_typeI}) (with $t_{k,m}$ ignored). These power allocation values are used in (\ref{compute X}) to compute $X_{k,m}$'s. After that, each subcarrier $k$ in the first time slot will independently select the subcarrier in the second time slot that gives the largest $X_{k,m}$ to maximize the the dual objective function (\ref{P_Lagrange_typeI}).

The last step is to find the values of
$\mu$ and ${\bm \alpha}$ which minimize $h(\mu,{\bm \alpha})$.
Using the subgradient method \cite{subgradient_method}, the values of $\mu$ and ${\bm \alpha}$ can be found iteratively as
\begin{equation}\label{sub_gradient_typeI_total}
\begin{aligned}
&\mu^{(i+1)}=\mu^{(i)}-y^{(i)}\left(P-\sum_{k=1}^M\sum_{m=1}^M~p^{(i)}_{k,m}\right),\\
&\alpha_m^{(i+1)}=\alpha_m^{(i)}-z^{(i)}\left(1-\sum_{k=1}^M~t_{k,m}^{(i)}\right)\text{, }m=1,...,M,
\end{aligned}
\end{equation}
where the superscript $(i)$ denotes the iteration index, and $y^{(i)}$ and $z^{(i)}$ are the sequences of step sizes designed
properly. With the new $\mu$ and ${\bm \alpha}$ in each iteration, the subcarrier pairing and power allocation can be updated with (\ref{optimal_t_typeI}) and (\ref{optimal power_typeI}), respectively, for the next iteration. As the number of iterations increases, (\ref{sub_gradient_typeI_total})
will converge to the dual optimum variables \cite{subgradient_method}. The optimal ${\bm \alpha}$, together with (\ref{optimal_t_typeI}), make $t_{k,m}^\ast$'s satisfy (\ref{constraint_t_k}) and (\ref{constraint_t_m}).


Note that with the optimal power allocation given in
(\ref{optimal power_typeI}), the achievable rate for SP $(k, m)$ is
\begin{equation}\label{impact_weighting}
\begin{aligned}
\frac{1}{2}\log\left(1+a_{k,m}\left[\frac{w_{k}}{2\mu}-\frac{1}{a_{k,m}}\right]^+\right)=\frac{1}{2}\log\left(1+a_{k,m}w_{k}\left[\frac{1}{2\mu}-\frac{1}{w_{k}a_{k,m}}\right]^+\right).
\end{aligned}
\end{equation}
From (\ref{impact_weighting}), the impact of the weighting factors can be viewed as weighting the channel gain of SP
$(k, m)$ by $w_{k}$. A higher weighting factor results in more power allocated to the corresponding SP.

%

\subsection{Discussion on the Duality Gap}\label{duality_gap_total}
\begin{figure}[htp]
\centering
\includegraphics[width=0.8\textwidth]{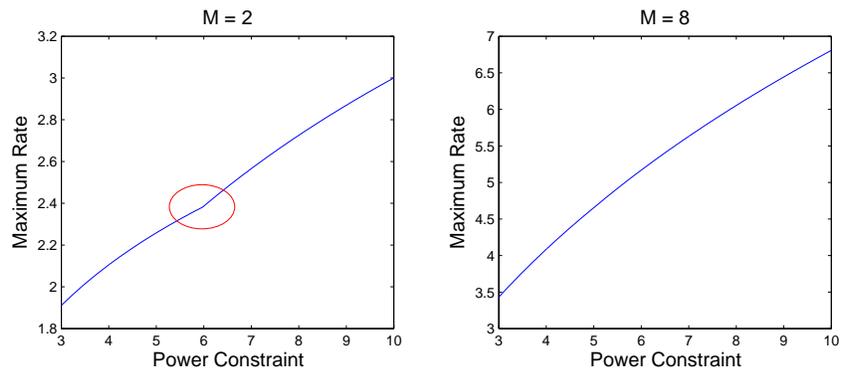}
\caption{Concavity of rate versus power constraint for systems with 2 and 8 subcarriers.}
\label{Fig_rate_power_concave}
\end{figure}

For problem (\ref{objective_Rate_I_MIP}), the optimal subcarrier pairing scheme may change as the total power constraint varies. Thus the maximum weighted sum rate as a function of the total power constraint may have discrete changes in the slope at the transition points where the optimal subcarrier pairing scheme changes. An example is shown in the circled region in Fig.~\ref{Fig_rate_power_concave} for $M=2$ subcarriers. This phenomenon is similar to that observed in the optimal resource allocation for OFDMA downlink systems \cite{zero_gap_SUB_infinity}. However, in our case, this phenomenon is observed even when the weighting factors for all subcarriers are set to the same. As discussed in \cite{Yu_dual}\cite{zero_gap_SUB_infinity}, the nonconcavity shown in Fig.~\ref{Fig_rate_power_concave} may result in nonzero duality gap.
Let us denote the optimal values of the original problem (\ref{objective_Rate_I_MIP}), the relaxed problem (\ref{objective_Rate_I_relaxed}), and the relaxed dual problem
(\ref{objective_dual}) by $R_B$, $R_R$, and $D_R$, respectively. The relationship between
them is $R_B\leq~R_R\leq~D_R$. Since the optimal $t_{k,m}^\ast$'s found by solving (\ref{P_Lagrange_typeI}) and (\ref{objective_dual}) satisfy (\ref{constraint_t_k}), (\ref{constraint_t_m}) and (\ref{constraint_t_binary}), we conclude
that $D_R$ is also the dual optimum value for problem
(\ref{objective_Rate_I_MIP}).

According to \cite{Yu_dual}\cite{Chiang_zero_gap}\cite{zero_gap_SUB_infinity}, the duality gap is zero if the optimal value of
the optimization problem is a concave function of the constraints. \cite{Yu_dual} and \cite{zero_gap_SUB_infinity} also showed analytically and through simulations that the concavity will be satisfied as the number of subcarriers becomes large.
In our case, we found that the concavity
is mostly satisfied when the number of subcarriers is reasonably large. Specifically, when $M=2$, we have observed in simulation that only about $1\%$ of the possible channel realizations will result in the nonconcavity shown in Fig.~\ref{Fig_rate_power_concave}. When $M=4$, the probability of nonconcavity is about $0.4\%$. For $M \geq 6$, the maximum weighted sum rate is almost always concave in the total power constraint. An example is shown in Fig.~\ref{Fig_rate_power_concave} for $M=8$ subcarriers. Thus, for practical OFDM systems, the duality gap is virtually zero, and $R_B\approx D_R$. We can then conclude that $R_B\approx R_R\approx D_R$ for most practical OFDM systems. This will be verified by the simulation results in Section~\ref{Sec_simulation}.

\subsection{Algorithm Design}\label{Sec_achieave_feasible}
Combining (\ref{optimal_t_typeI}), (\ref{optimal power_typeI}) and (\ref{sub_gradient_typeI_total}), the algorithm to find the optimal subcarrier pairing and power allocation can be designed as in the upper part of Table~\ref{Table_algo_obtain_feasible_typeI}.
However, through simulation, we have observed that although (\ref{optimal_t_typeI}) guarantees that each row of ${\bm t}$ has only one ``1", some of the ``1"s may be on the same column. This corresponds to the situation where more than one source subcarriers select the same relay subcarrier. As a result, the constraint (\ref{constraint_t_k}) is violated, and the solution is not feasible. This situation usually arises in two scenarios. The first is when there are more than one source subcarriers with very strong SD gains, such that no matter which relay subcarrier they are paired with, the direct-link mode will be selected. For any of these source subcarriers, the power related terms in $X_{k,m}$ (\ref{compute X}) are the same for all relay subcarriers. Thus the relay subcarrier selection (\ref{optimal_t_typeI}) depends only on $\alpha_m$. The source subcarriers with this property will select the same relay subcarrier. The other scenario is when a source subcarrier gets a low equivalent SP channel gain $a_{k,m}$ no matter which relay subcarrier it is paired with. All the possible SPs formed by this source subcarrier will be allocated very little power, thus their $X_{k,m}$'s are dominated by the corresponding $\alpha_m$'s. Similarly, the source subcarriers with this property will most likely select the same relay subcarrier.

To handle this situation, we include an amendment algorithm in the original algorithm as shown in the lower part of Table~\ref{Table_algo_obtain_feasible_typeI}. Based on the above discussion, the basic idea of the amendment algorithm is to each time move a ``1" in a column of ${\bm t}$ with more than one ``1"s to the column with no ``1" that will cause the minimum change in the value of $\alpha_m$.
%
By moving a ``1" to another column with a similar $\alpha_m$ value, the weighted sum rate will not be lowered much. When doing so, the amendment algorithm will make sure to keep the ``1" corresponding to the largest $X_{k,m}$ for each column with more than one ``1"s. It will also move the redundant ``1"s to the columns with no ``1" that will result in as large $X_{k,m}$ values as possible. Thus the resultant weighted sum rate will be maximized. Eventually the pairing scheme ${\bm t}$ altered by the amendment algorithm will meet the constraints (\ref{constraint_t_k}) and (\ref{constraint_t_m}).

The amendment algorithm is triggered when the dual variables converge to a certain degree (for the example in Table~\ref{Table_algo_obtain_feasible_typeI}, within 1\%). Once the amendment algorithm is triggered, the algorithm will continue to run for another 10\% of iterations. For example, if the amendment algorithm is triggered at the 1000th iteration, the algorithm will run another 100 iterations before it outputs the solution.
For each of these 10\% of iterations, a feasible pairing scheme will be obtained by the amendment algorithm. Using this pairing scheme, regular water-filling over parallel channels will be applied to obtain the optimal power allocation and the corresponding weighted sum rate. The best pairing scheme and power allocation among these iterations that achieve the highest weighted sum rate will be the outputs of this algorithm. As shown in Section \ref{Sec_simulation}, the weighted sum rate obtained by the algorithm in Table~\ref{Table_algo_obtain_feasible_typeI} is quite close to the optimal.

\subsection{Complexity Comparison}
The total number of all possible pairing schemes is $O(M!)$. With a fixed subcarrier pairing scheme, the complexity of computing the optimal power allocation (\ref{optimal power_typeI}) for the selected pairs is $O(M)$ in terms of multiplications. The complexity of computing the resulting weighted sum rate (weighted sum of (\ref{unified_rate})) is also $O(M)$ in terms of $\log(\cdot)$ operations and multiplications.
Thus the complexity of exhaustive search is $O(M\cdot M!)$ which is prohibitively high.

On the other hand, in
each iteration of the algorithm in Table~\ref{Table_algo_obtain_feasible_typeI}, the complexity is dominated by the computation of $X_{k,m}, \forall k,m$, in (\ref{compute X}). That complexity is $O(M^2)$ in terms of $\log(\cdot)$ operations and multiplications. For the amendment algorithm in the last 10\% of iterations, alteration of the pairing scheme ${\bm t}$ takes only $O(M^2)$ additions and $\max(\cdot)$ and $\min(\cdot)$ operations. The complexity of computing the optimal power allocation and the resulting weighted sum rate is $O(M)$ multiplications and $\log(\cdot)$ operations. Therefore the overall complexity for the algorithm in Table~\ref{Table_algo_obtain_feasible_typeI} is $O(JM^2)$, where $J$ is the number of iterations. This complexity is much more feasible and tractable.


\section{Weighted Sum Rate Maximization under Individual Power Constraints}\label{sec_maximization_individual}
When the source and the relay have individual power constraints, the weighted sum rate maximization problem becomes
\begin{align}\label{objective_Rate_I_individual}
\max_{{\bm p}_{S},{\bm p}_{R},{\cal S}_S,{\cal S}_R} & ~\frac{w_{k}}{2} \left(\sum_{(k,m) \in {\cal S}_S}\log\left(1+a^{SD}_{k}p^{S}_{k,m}\right) \right.\\
&\left.+\sum_{(k,m) \in {\cal S}_R} \min\left\{\log\left(1+a^{SR}_{k}p^{S}_{k,m}\right),\; \log\left(1+a^{SD}_{k}p^{S}_{k,m}+a^{RD}_{m}p^{R}_{k,m}\right)\right\} \right)
\\
\label{constraint_individual_power_0}
\text{s.t.}\text{ }&p^S_{k,m}, ~p^{R}_{k,m}\geq~0,\forall k, m
\\
\label{constraint_source_power_consumption}
&\sum_{(k,m) \in {\cal S}_S \cup {\cal S}_R} p^S_{k,m}\leq P_{S}
\\
\label{constraint_relay_power_consumption}
&\sum_{(k,m) \in {\cal S}_R} p^R_{k,m}\leq P_{R}
\end{align}
where $P_{S}$ and $P_{R}$ are the source and the
relay power constraints, respectively, and
${\bm p}_{S} \in \mathds{R}_+^{M\times M}$ and
${\bm p}_{R} \in \mathds{R}_+^{M\times M}$ are the
matrices of $p^S_{k,m}$ and $p^R_{k,m}$, respectively. ${\cal S}_S$ and ${\cal S}_R$ denote the sets of SPs operating in the direct-link mode and the relay mode, respectively. If we let $t_{k,m}=1$ when $(k,m) \in {\cal S}_S \cup {\cal S}_R$ and $t_{k,m}=0$ otherwise, ${\cal S}_S$ and ${\cal S}_R$ must satisfy the additional constraints (\ref{constraint_t_k}) and (\ref{constraint_t_m}).

This problem is very complicated. Because the condition to use relay depends not only on the channel condition, but also indirectly on the source power and relay power constraints \cite{DF_OFDM_total_individual_power}\cite{relay_DF_individual_power_constraint}\cite{Vandendorpe_J}, it is not possible to classify the SPs into the direct-link mode or the relay mode in advance to use the unified weighted rate formulation (\ref{unified_rate}) and the equivalent channel gain (\ref{Eq_ECG}). In Section~\ref{unified_rate_individual}, we will first investigate optimal power allocation with fixed subcarrier pairing under individual power constraints considered in
\cite{DF_OFDM_total_individual_power}\cite{relay_DF_individual_power_constraint}\cite{Vandendorpe_J}.
Through some insightful observations on the results of \cite{relay_DF_individual_power_constraint}, we will find that the unified weighted rate formulation (\ref{unified_rate}) and the equivalent channel gain (\ref{Eq_ECG}) can, in fact, be applied to the dual problem of (\ref{objective_Rate_I_individual}). After that, (\ref{objective_Rate_I_individual}) can be solved similarly as in the total power constrained case.

\subsection{Unified Rate Formulation}\label{unified_rate_individual}
For (\ref{objective_Rate_I_individual}), assuming fixed subcarrier pairing, a Lagrangian similar to \cite[eq. (8)]{relay_DF_individual_power_constraint} with slight changes can be obtained
\begin{equation}\label{Louveaux_Lagrange}
\begin{aligned}
L&=\sum_{(k,m) \in {\cal S}_S}\frac{w_{k}}{2}\log\left(1+a^{SD}_{k}p^{S}_{k,m}\right)\\
&+\sum_{(k,m) \in {\cal S}_R}\frac{w_{k}}{2}\log\left(1+a^{SD}_{k}p^{S}_{k,m}+a^{RD}_{m}p^{R}_{k,m}\right)\\
&+\mu_{S}\left(P_{S}-\sum_{k=1}^M p^{S}_{k,m}\right)+\mu_{R}\left(P_{R}-\sum_{(k,m) \in {\cal S}_R}
 p^{R}_{k,m}\right)\\
&+\sum_{(k,m) \in {\cal S}_R}\rho_{k,m}\left(a^{SR}_{k}p^{S}_{k,m}-a^{SD}_{k}p^{S}_{k,m}-a^{RD}_{m}p^{R}_{k,m}\right),
\end{aligned}
\end{equation}
where $\mu_S\geq 0$ and $\mu_R \geq 0$ denote the Lagrange multipliers corresponding to the source power constraint $P_S$ and the relay power constraint $P_R$, respectively. The third Lagrange multiplier $\rho_{k,m} \geq 0$ corresponds to the condition
\begin{equation}\label{Louveaux_relay_condition}
a^{SR}_{k}p^{S}_{k,m} \geq a^{SD}_{k}p^{S}_{k,m}+a^{RD}_{m}p^{R}_{k,m}
\end{equation}
for the SP $(k,m)$ to operate in the relay mode \cite{relay_DF_individual_power_constraint}\cite{Vandendorpe_J}. If (\ref{Louveaux_relay_condition}) is not valid for a relay mode SP, apparently some of that SP's relay power can be reallocated, without reducing its rate, to other relay subcarriers to improve their rates, or simply to be conserved. Following the same procedures as in \cite{relay_DF_individual_power_constraint}, a SP $(k,m)$ can be further
classified into the following three modes given that $\mu_S$ and $\mu_R$ are fixed:
\begin{equation}\label{mode_conditions_individual}
\left\{
\begin{aligned}
&\text{Direct-link mode}:&&a^{SD}_{k}\geq~a^{SR}_{k}\text{ or
}a^{RD}_{m}<~a^{SD}_{k}\frac{\mu_{R}}{\mu_{S}}\\
&\text{Relay mode}:&&a^{SR}_{k}>a^{SD}_{k}\text{ and } a^{RD}_{m}>a^{SD}_{k}\frac{\mu_{R}}{\mu_{S}}\\
&\text{Intermediate mode}:&&a^{SR}_{k}>a^{SD}_{k}\text{ and } a^{RD}_{m}=a^{SD}_{k}\frac{\mu_{R}}{\mu_{S}},\\
\end{aligned}
\right.
\end{equation}
where the intermediate mode is a special case of the relay mode with the condition (\ref{Louveaux_relay_condition}) satisfied with strict inequality (and the corresponding $\rho_{k,m}=0$). That is, the relay receives more
information than the destination. Thus the relay mode here is redefined to include only the SPs that satisfy (\ref{Louveaux_relay_condition}) with equality. That is, the amounts of received information are the same at the relay and the destination. According to \cite{relay_DF_individual_power_constraint}, usually there is at most one SP in the intermediate mode. For both the relay mode and the intermediate mode SPs, the solutions that maximize the Lagrangian (\ref{Louveaux_Lagrange}) will make its last term zero.

The first condition $a^{SD}_{k}\geq a^{SR}_{k}$ for selecting the direct-link mode over the relay mode for SP $(k,m)$ is based on the fact that, in this situation, the destination will receive more information than the relay. Then there is no need to use the relay. The second condition $a^{RD}_{m}<a^{SD}_{k}\frac{\mu_{R}}{\mu_{S}}$ ensures that the direct-link mode will contribute more to the Lagrangian (\ref{Louveaux_Lagrange}) than the relay mode. For example, for SP $(k,m)$, if the power ``cost" for selecting the direct-link mode, $\mu_S p^{S}_{k,m}$, in (\ref{Louveaux_Lagrange}) is kept the same as the power cost for selecting the relay mode, $\mu_S p^{S}_{k,m}+\mu_R p^{R}_{k,m}$, when $a^{RD}_{m}<a^{SD}_{k}\frac{\mu_{R}}{\mu_{S}}$, selecting the direct-link mode will result in a higher weighted rate (hence larger value of (\ref{Louveaux_Lagrange})) than selecting the relay mode. This is implied by the ``single sum power constraint'' approach in \cite[eqs. (39) and (40)]{Vandendorpe_J} with modified power and channel gain variables.

By solving (\ref{Louveaux_Lagrange}) for weighted sum rate maximization, the power allocation can be obtained
\begin{equation}\label{Louveaux_power_allocation}
\left\{
\begin{aligned}
&\text{Direct-link mode}:&& p^{S}_{k,m}=\left[\frac{w_{k}}{2\mu_{S}}-\frac{1}{a^{SD}_{k}}\right]^+, ~p^{R}_{k,m}=0\\
&\text{Relay mode}:&& p^{S}_{k,m}=\left[\frac{w_{k}}{2\left(\mu_{S}+\mu_{R}/\beta_{k,m}\right)}-\frac{1}{a^{SR}_{k}}\right]^+, ~p^{R}_{k,m}=\left[\frac{w_{k}}{2\left(\mu_{S}\beta_{k,m}+\mu_{R}\right)}-\frac{1}{a^{SR}_{k}\beta_{k,m}}\right]^+\\
&\text{Intermediate mode}:&& \text{According to } ~1+a^{SD}_{k}p^{S}_{k,m}+a^{RD}_{m}p^{R}_{k,m}=\frac{w_{k}a^{SD}_{k}}{2\mu_{S}}, ~~~a^{RD}_{m}=a^{SD}_{k}\frac{\mu_{R}}{\mu_{S}},\\
&&&\text{and source and relay power constraints}, \\
\end{aligned}
\right.
\end{equation}
where
\begin{equation}
\beta_{k,m}=\frac{a^{RD}_m}{a^{SR}_k-a^{SD}_k}.
\end{equation}
Power allocation for the intermediate mode SP can be computed after the power allocations for the direct-link mode and relay mode SPs are done. Note that since the relay mode SPs must satisfy (\ref{Louveaux_relay_condition}) with equality, it is clear that $p^S_{k,m}=\beta_{k,m}p^R_{k,m}$. From (\ref{mode_conditions_individual}) we know that for the relay mode, $\beta_{k,m}>0$. Thus $p^{S}_{k,m}$ and $p^{R}_{k,m}$ must be zero or positive simultaneously. This can also be seen from the relay mode power allocation in (\ref{Louveaux_power_allocation}). This observation allows us to allocate total power $p_{k,m}=p^{S}_{k,m}+ p^{R}_{k,m}$ to the relay mode SPs first according to
\begin{equation}\label{power_allocation_relay_individual}
p_{k,m}=\left(\beta_{k,m}+1\right)\left[\frac{w_{k}}{2\left(\mu_{S}\beta_{k,m}+\mu_{R}\right)}-\frac{1}{a^{SR}_{k}\beta_{k,m}}\right]^+,
\end{equation}
then obtain the corresponding $p^{S}_{k,m}$ and $p^{R}_{k,m}$ using the relay mode power distribution in (\ref{power_ratio}). To this end, the unified weighted rate expression in (\ref{unified_rate}) with (\ref{power_ratio}) and the equivalent channel gain (\ref{Eq_ECG}) can be applied here as well to the direct-link mode and relay mode SPs, when $\mu_S$ and $\mu_R$ are fixed.

As to the intermediate mode SP, we examine its contributions to the rate and cost in the Lagrangian (\ref{Louveaux_Lagrange}) and find that, with $\rho_{k,m}=0$ \cite{relay_DF_individual_power_constraint}, they are
\begin{equation}
\begin{aligned}
\text{rate}&=\frac{w_{k}}{2}\log\left(1+a^{SD}_{k}p^{S}_{k,m}+a^{RD}_{m}p^{R}_{k,m}\right)=\frac{w_{k}}{2}\log\left(\frac{w_{k}a^{SD}_{k}}{2\mu_{S}}\right)\\
\text{cost}&=\mu_{S}p^{S}_{k,m}+\mu_{R}p^{R}_{k,m}=\mu_{S}\left(p^{S}_{k,m}+\frac{\mu_{R}}{\mu_{S}}p^{R}_{k,m}\right)\\
&=\mu_{S}\left(p^{S}_{k,m}+\frac{a^{RD}_{m}}{a^{SD}_{k}}p^{R}_{k,m}\right)=\frac{\mu_{S}}{a^{SD}_{k}}\left(a^{SD}_{k}p^{S}_{k,m}+a^{RD}_{m}p^{R}_{k,m}\right)\\
&=\frac{\mu_{S}}{a^{SD}_{k}}\left(\frac{w_{k}a^{SD}_{k}}{2\mu_{S}}-1\right)=\mu_{S}\left(\frac{w_{k}}{2\mu_{S}}-\frac{1}{a^{SD}_{k}}\right)
\end{aligned}.
\end{equation}
If the intermediate mode SP was classified as a direct-link mode SP, it would contribute to the Lagrangian (\ref{Louveaux_Lagrange}) with
\begin{equation}
\begin{aligned}
\text{rate}&=\frac{w_{k}}{2}\log\left(1+a^{SD}_{k}p^{S}_{k,m}\right)=\frac{w_{k}}{2}\log\left(1+a^{SD}_{k}\left(\frac{w_{k}}{2\mu_{S}}-\frac{1}{a^{SD}_{k}}\right)\right)\\
&=\frac{w_{k}}{2}\log\left(\frac{w_{k}a^{SD}_{k}}{2\mu_{S}}\right)\\
\text{cost}&=\mu_{S}p^{S}_{k,m}=\mu_{S}\left(\frac{w_{k}}{2\mu_{S}}-\frac{1}{a^{SD}_{k}}\right)
\end{aligned}.
\end{equation}
On the other hand, if it was classified as a relay mode SP, with $\rho_{k,m}=\frac{a^{RD}_{m}\mu_S-a^{SD}_{k}\mu_R}{a^{SR}_{k}a^{RD}_{m}}=0$, its contributions to the Lagrangian (\ref{Louveaux_Lagrange}) would be
\begin{equation}
\begin{aligned}
\text{rate}&=\frac{w_{k}}{2}\log\left(1+a^{SD}_{k}p^{S}_{k,m}+a^{RD}_{m}p^{R}_{k,m}\right)\\
&=\frac{w_{k}}{2}\log\left(1+\frac{w_{k}a^{SD}_{k}}{2\left(\mu_{S}+\mu_{R}/\beta_{k,m}\right)}-\frac{a^{SD}_{k}}{a^{SR}_{k}}+\frac{w_{k}a^{RD}_{m}/\beta_{k,m}}{2\left(\mu_{S}+\mu_{R}/\beta_{k,m}\right)}-\frac{a^{RD}_{m}/\beta_{k,m}}{a^{SR}_{k}}\right)\\
&=\frac{w_{k}}{2}\log\left(\frac{w_{k}a^{SR}_{k}}{2\left(\mu_{S}+\mu_{S}\frac{a^{RD}_{m}}{a^{SD}_{k}\beta_{k,m}}\right)}\right)=\frac{w_{k}}{2}\log\left(\frac{w_{k}a^{SD}_{k}}{2\mu_{S}}\right)\\
\text{cost}&=\mu_{S}p^{S}_{k,m}+\mu_{R}p^{R}_{k,m}=\frac{\mu_{S}}{a^{SD}_{k}}\left(a^{SD}_{k}p^{S}_{k,m}+a^{RD}_{m}p^{R}_{k,m}\right)=\frac{\mu_{S}}{a^{SD}_{k}}\left(a^{SR}_{k}p^{S}_{k,m}\right)\\
&=\frac{\mu_{S}a^{SR}_{k}}{a^{SD}_{k}}\left(\frac{w_{k}}{2\left(\mu_{S}+\mu_{R}/\beta_{k,m}\right)}-\frac{1}{a^{SR}_{k}}\right)\\
&=\mu_{S}\left(\frac{w_{k}}{2\mu_{S}}-\frac{1}{a^{SD}_{k}}\right)
\end{aligned}.
\end{equation}
Interestingly, with given $\mu_S$ and $\mu_R$, the intermediate mode SP's contributions to the Lagrangian (\ref{Louveaux_Lagrange}) remain the same no matter it is classified to the direct-link mode or the relay mode. Thus, in terms of maximizing the Lagrangian, we can assign the intermediate mode SP to either mode without affecting the result. However, once the optimal $\mu_S$ and $\mu_R$ are obtained, we still need to identify the intermediate mode SP and allocate its powers according to (\ref{Louveaux_power_allocation}).

In the following, we will assign the intermediate mode SP to the relay mode. Together with the conclusion that the unified weighted rate and equivalent channel gain expressions can be applied when $\mu_S$ and $\mu_R$ are fixed, the dual problem of (\ref{objective_Rate_I_individual}) can be formulated with unified expressions.

\subsection{Dual Problem}\label{sec_dual_type_I_individual}
By dualizing (\ref{constraint_t_k}), (\ref{constraint_source_power_consumption}), (\ref{constraint_relay_power_consumption}), (\ref{Louveaux_relay_condition}), letting ${\bm t}$ and ${\bm \rho}$ be the matrices of $t_{k,m}$ and $\rho_{k,m}$, respectively, and applying continuous relaxation to $t_{k,m}$'s as in Section~\ref{primal_total}, we have the following Lagrangian
\begin{equation}\label{Lagrangian_individual_OK_typeI}
\begin{aligned}
L({\bm p},{\bm t},\mu_{S},\mu_{R}, {\bm \alpha}, {\bm \rho})=&\sum_{k=1}^M\sum_{m=1}^M~t_{k,m}\frac{w_{k}}{2}\log\left(1+a_{k,m}\frac{p_{k,m}}{t_{k,m}}\right)\\
&+\mu_{S}\left(P_{S}-\sum_{k=1}^M\sum_{m=1}^M~c^{S}_{k,m}~p_{k,m}\right)\\
&+\mu_{R}\left(P_{R}-\sum_{k=1}^M\sum_{m=1}^M~c^{R}_{k,m}~p_{k,m}\right)\\
&+\sum_{m=1}^M\alpha_m\left(1-\sum_{k=1}^M~t_{k,m}\right)\\
&+\sum_{(k,m) \in {\cal S}_R}\rho_{k,m}\left(a^{SR}_{k}p^{S}_{k,m}-a^{SD}_{k}p^{S}_{k,m}-a^{RD}_{m}p^{R}_{k,m}\right),
\end{aligned}
\end{equation}
where
\begin{equation}\label{Eq_ECG_individual}
a_{k,m}=\left\{
\begin{aligned}
&\frac{a^{SR}_{k}a^{RD}_{m}}{a^{SR}_{k}+a^{RD}_{m}-a^{SD}_{k}},&&\text{when }a^{SR}_{k}>a^{SD}_{k} \text{ and } a^{RD}_{m}\geq~a^{SD}_{k}\frac{\mu_{R}}{\mu_{S}}\\
&a^{SD}_{k},&&\,\,\,\text{otherwise}
\end{aligned}
\right.
\end{equation}
\begin{equation}\label{source_power_ratio_individual}
c^{S}_{k,m}=\left\{
\begin{aligned}
&\frac{a^{RD}_{m}}{a^{SR}_{k}+a^{RD}_{m}-a^{SD}_{k}},&&\text{when }a^{SR}_{k}>a^{SD}_{k} \text{ and } a^{RD}_{m}\geq~a^{SD}_{k}\frac{\mu_{R}}{\mu_{S}}\\
&1,&&\,\,\,\text{otherwise}
\end{aligned}
\right.
\end{equation}
\begin{equation}\label{relay_power_ratio_individual}
c^{R}_{k,m}=\left\{
\begin{aligned}
&\frac{a^{SR}_{k}-a^{SD}_{k}}{a^{SR}_{k}+a^{RD}_{m}-a^{SD}_{k}},&&\text{when }a^{SR}_{k}>a^{SD}_{k} \text{ and } a^{RD}_{m}\geq~a^{SD}_{k}\frac{\mu_{R}}{\mu_{S}}\\
&0,&&\,\,\,\text{otherwise}
\end{aligned}
\right.
\end{equation}
are the equivalent channel gain, the portions of $p_{k,m}$
distributed to source power and relay power, respectively, for
the two modes specified in the conditions.
Similar to (\ref{objective_dual}), the dual problem associated with (\ref{Lagrangian_individual_OK_typeI}) can be expressed as
\begin{equation} \label{objective_dual_typeI_indivudual}
\min_{\mu_{S},\mu_{R}, {\bm \alpha}} ~h(\mu_{S},\mu_{R}, {\bm \alpha})
~~\text{s.t.}~~ \mu_{S}\geq 0, ~\mu_{R}\geq 0
\end{equation}
with
\begin{equation}\label{P_Lagrange_typeI_indivudial}
h(\mu_{S},\mu_{R}, {\bm \alpha})=
\max_{{\bm p},{\bm t},{\bm \rho}} ~L({\bm p},{\bm t},\mu_{S},\mu_{R}, {\bm \alpha}, {\bm \rho})
~~\text{s.t.}~~ (\ref{constraint_t_m}), (\ref{constraint_power_0}), (\ref{constraint_t_0}).
\end{equation}
Note that the source and relay power distribution (\ref{source_power_ratio_individual}), (\ref{relay_power_ratio_individual}) satisfy the constraint $a^{SR}_{k}p^{S}_{k,m} = a^{SD}_{k}p^{S}_{k,m}+a^{RD}_{m}p^{R}_{k,m}$ for the relay mode. Thus the last term in (\ref{Lagrangian_individual_OK_typeI}) is always zero. Following the same procedure as in Section~\ref{sec_dual_type_I_total} and applying the results in Section~\ref{unified_rate_individual}, the optimal power allocation for (\ref{P_Lagrange_typeI_indivudial}) can be solved as
\begin{equation}
\label{optimal power_typeI_individual}
p_{k,m}^\ast=t_{k,m}\left[\frac{w_{k}}{2(c^{S}_{k,m}\mu_{S}+c^{R}_{k,m}\mu_{R})}-\frac{1}{a_{k,m}}\right]^+.
\end{equation}
The optimal $t_{k,m}$ can be solved as
\begin{align}
\label{optimal_t_typeI_individual}
t_{k,m}^\ast=&\left\{
\begin{aligned}
&1,\text{ }m=\arg\max_{m=1,...,M}Z_{k,m}\\
&0,\text{ otherwise}
\end{aligned}
\right., ~~~~\forall k.
\end{align}
where
\begin{equation}
\begin{aligned}\label{compute Z}
Z_{k,m}=&\frac{w_{k}}{2}\log\left(1+a_{k,m}\left[\frac{w_{k}}{2(c^{S}_{k,m}\mu_{S}+c^{R}_{k,m}\mu_{R})}-\frac{1}{a_{k,m}}\right]^+\right)-\alpha_m\\
&-\mu_{S}\left(c^{S}_{k,m}\left[\frac{w_{k}}{2(c^{S}_{k,m}\mu_{S}+c^{R}_{k,m}\mu_{R})}-\frac{1}{a_{k,m}}\right]^+\right)\\
&-\mu_{R}\left(c^{R}_{k,m}\left[\frac{w_{k}}{2(c^{S}_{k,m}\mu_{S}+c^{R}_{k,m}\mu_{R})}-\frac{1}{a_{k,m}}\right]^+\right).
\end{aligned}
\end{equation}

Since assigning the intermediate mode SP to the relay mode does not change the dual value, we can approach the dual optimal value by the subgradient method.
The Lagrange multipliers $\mu_{S}$, $\mu_{R}$, and ${\bm \alpha}$ are updated by
\begin{equation}\label{sub_gradient_typeI individual}
\begin{aligned}
&\mu_{S}^{(i+1)}=\mu_{S}^{(i)}-y_{S}^{(i)}\left(P_{S}-\sum_{k=1}^M\sum_{m=1}^M~c^{S}_{k,m}p_{k,m}^{(i)}\right),\\
&\mu_{R}^{(i+1)}=\mu_{R}^{(i)}-y_{R}^{(i)}\left(P_{R}-\sum_{k=1}^M\sum_{m=1}^M~c^{R}_{k,m}p_{k,m}^{(i)}\right),\\
&\alpha_m^{(i+1)}=\alpha_m^{(i)}-z^{(i)}\left(1-\sum_{k=1}^M~t_{k,m}^{(i)}\right)\text{, }m=1,...,M,
\end{aligned}
\end{equation}
where $y^{(i)}_S$, $y^{(i)}_R$ and $z^{(i)}$ are the sequences of step sizes designed properly. When the optimal subcarrier pairing and power allocation do not include an intermediate mode SP, (\ref{sub_gradient_typeI individual}) will converge to the optimal values. However, when an intermediate mode SP is present in the optimal solution, (\ref{sub_gradient_typeI individual}) may oscillate around the optimal values. Specifically, due to assigning the intermediate mode SP to the relay mode with power allocation (\ref{optimal power_typeI_individual}), the relay power for that SP is increased, while the source power is decreased, to make (\ref{Louveaux_relay_condition}) satisfied with equality instead of strict inequality. Thus, even when $\mu_S$ and $\mu_R$ are already at their optimal, the total source power consumption will be smaller than the source power constraint, and the total relay power consumption will be larger than the relay power constraint. This will result in $\mu_S$ decreased and $\mu_R$ increased in the next iteration. Then $\mu_R/\mu_S$ will be increased, and the intermediate mode SP may fall in the direct-link mode according to (\ref{mode_conditions_individual}). Similarly, this will make $\mu_R/\mu_S$ decreased, and the intermediate mode SP may fall in the relay mode in the next iteration. As a result, (\ref{sub_gradient_typeI individual}) oscillates. Similar oscillation was also observed in \cite{relay_DF_individual_power_constraint}. Thus, like in \cite[Section 3.2]{relay_DF_individual_power_constraint}, the zero-crossing of the difference between the total source power consumption and the source power constraint can be used to determine the optimal $\mu_R/\mu_S$ and the corresponding mode classification and power allocation. However, due to the issues discussed in Section \ref{Sec_achieave_feasible}, we have found that the optimal zero-crossing is very difficult to trace when subcarrier pairing, mode classification and power allocation are updated at the same time. In the algorithm given in Table~\ref{Table_algo_obtain_feasible_typeI_individual}, similar to Table~\ref{Table_algo_obtain_feasible_typeI}, the amendment algorithm is used to obtain a feasible pairing scheme when the subgradient method converges to a ceratin degree. With diminishing step sizes, we found that the subgradient method will eventually be stuck at assigning the intermediate mode SP (if it exists in the optimal solution) to either the direct-link mode or the relay mode. In both cases, the obtained subcarrier pairing scheme is near optimal. With fixed subcarrier pairing, and $\mu_R/\mu_S$ given by the amendment algorithm which is already very close to the optimal, the zero-crossing method in \cite[Section 3.2]{relay_DF_individual_power_constraint} can be used to quickly obtain the optimal $\mu_R/\mu_S$. Then the corresponding mode classification and power allocation can be done according to (\ref{mode_conditions_individual}) and (\ref{Louveaux_power_allocation}), respectively.

The algorithm in Table~\ref{Table_algo_obtain_feasible_typeI_individual} has the same order of complexity as that of the algorithm in Table~\ref{Table_algo_obtain_feasible_typeI}. Through simulation, we have also found that the duality gap for this problem approaches zero when the number of subcarriers is reasonably large.

\section{Weighted Sum Rate Maximization with Extra Direct-Link Transmission}\label{sec_maximization_total_extra_direct}
In the previous sections, only the relay can transmit in the second time slot.
Therefore, for the SPs operating in the direct-link mode, the second time slot is not used. It is possible to allow the source to transmit extra messages in the second time slot on these idle subcarriers. We consider this modified system with both total and individual power constraints.

\subsection{Total Power Constraint}\label{sec_extraSD_total}
Under the total power constraint, the achievable weighted sum rate for SP $(k,m)$ for this system is
\begin{equation}\label{Eq_pair_rate_typeII}
R_{k,m}=\left\{
\begin{aligned}
&\frac{w_{k}}{2}\log\left(1+a^{SD}_{k}p^{S}_{k,m}\right)+\frac{w_{m}}{2}\log\left(1+a^{SD}_{m}q^{S}_{k,m}\right),&\text{direct-link mode}\\
&\frac{w_{k}}{2}\min\left\{\log\left(1+a^{SR}_{k}p^{S}_{k,m}\right),\; \log\left(1+a^{SD}_{k}p^{S}_{k,m}+a^{RD}_{m}p^{R}_{k,m}\right)\right\}, &\text{relay mode},\\
\end{aligned}
\right.
\end{equation}
where $p^{S}_{k,m}$, $p^{R}_{k,m}$, and $q^{S}_{k,m}$ represent the
source power in the first time slot, relay power in the second
time slot, and source power in the second time slot, respectively.
By comparing the achievable weighted rate for these two modes, we find that the condition for using the relay depends not only on the channel gains but also on the power allocation. Thus we introduce an additional indicator $s_{k,m}$ related to the use of the relay as a variable to be jointly optimized. When $s_{k,m}=1$, the relay is used for SP $(k,m)$. When $s_{k,m}=0$, the relay is not used.
In addition, we again make continuous relaxation for the indicators and the same adjustment to the sum rate function. The relaxed weighted sum rate maximization problem is expressed as follows
\begin{align}
\label{objective_Rate_II_relaxed}
\max_{{\bm p},{\bm s},{\bm t}}\text{ }
&\frac{1}{2}\sum^M_{k=1}\sum^M_{m=1}t_{k,m}~\left\{s_{k,m}w_{k}\log\left(1+a_{k,m}\frac{p_{k,m,1}}{t_{k,m}s_{k,m}}\right)\right.\notag\\
&+(1-s_{k,m})\left[w_{k}\log\left(1+a^{SD}_{k}\frac{p_{k,m,2}}{t_{k,m}(1-s_{k,m})}\right)\right.\notag\\
&\left.\left.+w_{m}\log\left(1+a^{SD}_{m}\frac{p_{k,m,3}}{t_{k,m}(1-s_{k,m})}\right)\right]\right\}
\\
\text{s.t.}\text{ }&(\ref{constraint_t_k}), (\ref{constraint_t_m}), (\ref{constraint_t_0})
\notag\\
\label{constraint_power_k,m,r}
&\sum^M_{k=1}\sum^M_{m=1}\sum^3_{r=1}p_{k,m,r}\leq P
\\
\label{constraint_power_k,m,r_0}
&p_{k,m,r}\geq 0,\forall k, m, r
\\
\label{constraint_s_1_0}
&0\leq s_{k,m}\leq 1,\forall k, m,
\end{align}
where $p_{k,m,1}$ and $a_{k,m}$ are the sum power and equivalent channel gain, respectively, of the relay mode SP $(k,m)$ taking the form of the relay mode expressions in (\ref{power_ratio}) and (\ref{Eq_ECG}). $p_{k,m,2}$ and $p_{k,m,3}$ are the powers used by
the direct-link mode SP $(k,m)$ in the
first and second time slots, respectively. ${\bm p}\in \mathds{R}_+^{M\times M\times3}$,
${\bm t}\in \mathds{R}_+^{M\times M}$, and ${\bm s}\in \mathds{R}_+^{M\times M}$ are the
matrices of $p_{k,m,r}$, $t_{k,m}$, and $s_{k,m}$, respectively. $P$ is the total power constraint.

Similarly, by dualizing constraints (\ref{constraint_t_k})
and (\ref{constraint_power_k,m,r}), we obtain the Lagrangian as
\begin{equation}\label{Lagrangian_typeII}
\begin{aligned}
L({\bm p},{\bm t},{\bm s},\mu, {\bm \alpha})=&\frac{1}{2}\sum^M_{k=1}\sum^M_{m=1}t_{k,m}~\left\{s_{k,m}w_{k}\log\left(1+a_{k,m}\frac{p_{k,m,1}}{t_{k,m}s_{k,m}}\right)\right.\\
&+(1-s_{k,m})\left[w_{k}\log\left(1+a^{SD}_{k}\frac{p_{k,m,2}}{t_{k,m}(1-s_{k,m})}\right)\right.\\
&\left.\left.+w_{m}\log\left(1+a^{SD}_{m}\frac{p_{k,m,3}}{t_{k,m}(1-s_{k,m})}\right)\right]\right\}\\
&+\mu\left(P-\sum_{k=1}^M\sum_{m=1}^M\sum_{r=1}^3~p_{k,m,r}\right)+\sum_{m=1}^M\alpha_m\left(1-\sum_{k=1}^M~t_{k,m}\right),
\end{aligned}
\end{equation}
where $\mu\in \mathds{R}_+$ and ${\bm \alpha}\in\mathds{R}^M$ are the dual variables.
Then the dual objective function is computed as
\begin{equation} \label{P_Lagrange_typeII}
h(\mu,{\bm \alpha})=
\max_{{\bm p},{\bm t},{\bm s}} ~L({\bm p},{\bm t},{\bm s},\mu,{\bm \alpha})
~~\text{s.t.}~~ (\ref{constraint_t_m}), (\ref{constraint_t_0}), (\ref{constraint_power_k,m,r_0}), (\ref{constraint_s_1_0}).
\end{equation}
The dual problem is given as
\begin{equation} \label{objective_dual_typeII}
\min_{\mu,{\bm \alpha}} ~h(\mu, {\bm \alpha})
~~\text{s.t.}~~ \mu\geq 0.
\end{equation}
The solution to (\ref{P_Lagrange_typeII}) is
\begin{equation}\label{optimal power_typeII}
\begin{aligned}
&p_{k,m,1}^\ast=t_{k,m}s_{k,m}\left[\frac{w_{k}}{2\mu}-\frac{1}{a_{k,m}}\right]^+,\\
&p_{k,m,2}^\ast=t_{k,m}(1-s_{k,m})\left[\frac{w_{k}}{2\mu}-\frac{1}{a^{SD}_{k}}\right]^+,\\
&p_{k,m,3}^\ast=t_{k,m}(1-s_{k,m})\left[\frac{w_{m}}{2\mu}-\frac{1}{a^{SD}_{m}}\right]^+,
\end{aligned}
\end{equation}
\begin{equation}\label{Eq_s_choose}
s^\ast_{k,m}=\left\{
\begin{aligned}
1&,\text{ }a^{SR}_{k}>a^{SD}_{k}\text{ and }Y^{R}_{k,m}>Y^{D}_{k,m}\\
0&,\text{ otherwise},
\end{aligned}
\right.
\end{equation}
where
\begin{equation}\label{compute Y_R}
Y^{R}_{k,m}=\frac{w_{k}}{2}\log\left(1+a_{k,m}\left[\frac{w_{k}}{2\mu}-\frac{1}{a_{k,m}}\right]^+\right)-\mu\left[\frac{w_{k}}{2\mu}-\frac{1}{a_{k,m}}\right]^+\\
\end{equation}
\begin{equation}\label{compute Y_D}
\begin{aligned}
Y^{D}_{k,m}=&\frac{w_{k}}{2}\log\left(1+a^{SD}_{k}\left[\frac{w_{k}}{2\mu}-\frac{1}{a^{SD}_{k}}\right]^+\right)\\
&+\frac{w_{m}}{2}\log\left(1+a^{SD}_{m}\left[\frac{w_{m}}{2\mu}-\frac{1}{a^{SD}_{m}}\right]^+\right)\\
&-\mu\left(\left[\frac{w_{k}}{2\mu}-\frac{1}{a^{SD}_{k}}\right]^+~+\left[\frac{w_{m}}{2\mu}-\frac{1}{a^{SD}_{m}}\right]^+\right)\\
\end{aligned}
\end{equation}
are the SP $(k,m)$'s contribution to the Lagrangian when it is in the relay mode or the direct-link mode, respectively. The condition $a^{SR}_{k}>a^{SD}_{k}$ in (\ref{Eq_s_choose}) is
necessary. The reason is that the value of $Y^{R}_{k,m}$ is
meaningless when $a^{SR}_{k}<a^{SD}_{k}$, since it is impossible to
make the relay receive more information than the destination.
The $s^\ast_{k,m}$ tells us whether it is better to use relay for the SP ($k,m$).

The SP selection variable is given as follows
\begin{equation}\label{optimal_t_typeII}
t_{k,m}^\ast=\left\{
\begin{aligned}
&1,\text{ }m=\arg\max_{m=1,...,M}Y_{k,m}\\
&0,\text{ }\text{otherwise}
\end{aligned}
\right., ~~~~\forall k
\end{equation}
where
\begin{equation}\label{compute Y}
\begin{aligned}
Y_{k,m}=s^\ast_{k,m}Y^{R}_{k,m}+(1-s^\ast_{k,m})Y^{D}_{k,m}-\alpha_m.
\end{aligned}
\end{equation}
Again, the dual optimal value is reached by the subgradient method.
The Lagrange multipliers $\mu$ and ${\bm \alpha}$ are updated by
\begin{equation}\label{sub_gradient_typeII_total}
\begin{aligned}
&\mu^{(i+1)}=\mu^{(i)}-y^{(i)}\left(P-\sum_{k=1}^M\sum_{m=1}^M\sum_{r=1}^3~p^{(i)}_{k,m,r}\right),\\
&\alpha_m^{(i+1)}=\alpha_m^{(i)}-z^{(i)}\left(1-\sum_{k=1}^M~t_{k,m}^{(i)}\right)\text{, }m=1,...,M,
\end{aligned}
\end{equation}
where $y^{(i)}$ and $z^{(i)}$ are the sequences of step sizes designed properly.

The algorithm to obtain feasible solutions is given in Table \ref{Table_algo_obtain_feasible_typeII} where $s_{k,m}$'s found in an iteration are directly used, together with the subcarrier pairing scheme ${\bm t}$ obtained by the amendment algorithm, to compute the power allocation and weighted sum rate. Doing so is suboptimal, as $s_{k,m}$ in fact depends on the power allocation. However, this saves the complexity involved in joint optimization of $s_{k,m}$ and power allocation given fixed subcarrier pairing.
The algorithm in Table~\ref{Table_algo_obtain_feasible_typeII} also has the same order of complexity as that of the algorithm in Table~\ref{Table_algo_obtain_feasible_typeI}. We have found that the duality gap for this problem is virtually zero when the number of subcarriers is reasonably large.

\subsection{Individual Power Constraints}\label{sec_extraSD_individual}
With individual power constraints for the source and the relay, the problem can be solved by combining the results in Section~\ref{sec_maximization_individual} and Section~\ref{sec_extraSD_total} with some crucial modifications. Due to limited space, we will discuss only these crucial points.

As in Section~\ref{sec_maximization_individual}, in addition to the direct-link mode and the relay mode, there may also be an intermediate mode in which the relay receives more information than the destination, and (\ref{Louveaux_relay_condition}) is satisfied with strict inequality. On the other hand, the relay mode should satisfy (\ref{Louveaux_relay_condition}) with equality. Note that given the same power, the rate of the direct-link mode in (\ref{Eq_pair_rate_typeII}) with extra second-slot SD transmission should be no less than the rate of the direct-link mode in (\ref{Eq_pair_rate_typeI}). This is because the latter is a special case of the former with the second time slot allocated zero power. Therefore, the necessary condition  $a^{SR}_{k}>a^{SD}_{k}$ in (\ref{mode_conditions_individual}) for the relay to be active (in both the relay and the intermediate modes) is also necessary in this case. The second necessary condition $a^{RD}_{m}=a^{SD}_{k}\frac{\mu_{R}}{\mu_{S}}$ in (\ref{mode_conditions_individual}) for a SP to be in the intermediate mode, as derived in \cite{relay_DF_individual_power_constraint}, is directly related to having (\ref{Louveaux_relay_condition}) as a strict inequality. Thus it is also necessary in this case. The second condition $a^{RD}_{m}<a^{SD}_{k}\frac{\mu_{R}}{\mu_{S}}$ in (\ref{mode_conditions_individual}) for selecting the direct-link mode over the relay mode was derived by comparing the achievable rates of the direct-link mode and the relay mode when they have the same power cost in the Lagragian (\ref{Louveaux_Lagrange}) (see the discussion after (\ref{mode_conditions_individual}) and \cite{Vandendorpe_J}). With the extra second-slot SD transmission that can improve the rate for the direct-link mode, this condition may change. In fact, with the extra second-slot SD transmission, the direct-link mode may possibly be selected even when $a^{RD}_{m}>a^{SD}_{k}\frac{\mu_{R}}{\mu_{S}}$. For the special case with fixed $(k,k)$ subcarrier paring, \cite{Vandendorpe_J} has derived the exact condition which also depends on the allocated power. In our case, the subcarrier paring is variable and may not be the trivial $(k,k)$ pairing. Due to this reason and different weighting factors in the rate of the direct-link mode (\ref{Eq_pair_rate_typeII}), the exact condition based on having the same power cost is complicated and dependent also on the weighting factors. However, we may simplify the condition by comparing the contributions of the direct-link mode and the relay mode to the Lagrangian. This approach is similar to using (\ref{Eq_s_choose}) to select modes to maximize the Lagrangian (\ref{P_Lagrange_typeII}).

The fact that the direct-link mode may also be selected when $a^{RD}_{m}>a^{SD}_{k}\frac{\mu_{R}}{\mu_{S}}$ implies that the region for the intermediate mode to occur may be encompassed by the region for selecting the direct-link mode. That is, the intermediate mode may no longer exist, except in the special situation where the optimal power allocation for a direct-link mode SP results in zero power for the second-slot SD transmission. For a SP with this property, there will be no second-slot SD transmission if the direct-link mode is selected. Then the situation becomes the same as in Section~\ref{sec_maximization_individual}. Thus (\ref{mode_conditions_individual}) can be used to select modes, and the unified rate formulation discussed in Section~\ref{unified_rate_individual} can be applied with the intermediate mode SP assigned to the relay mode.

In summary, we can assume that there are only the direct-link mode and the relay mode, and apply the results in Section~\ref{sec_extraSD_total} with the following changes. The relay mode power allocation $p_{k,m,1}^\ast$ in (\ref{optimal power_typeII}) takes the expression of the right-hand side (RHS) of (\ref{optimal power_typeI_individual}) multiplied by $s_{k,m}$, where $c_{k,m}^S$ and $c_{k,m}^R$ are the portions of $p_{k,m,1}^\ast$ distributed to source power and relay power defined by the relay mode expressions in (\ref{source_power_ratio_individual}) and (\ref{relay_power_ratio_individual}), respectively. The RHS of (\ref{compute Y_R}) is replaced by the RHS of (\ref{compute Z}) with $-\alpha_m$ removed. $\mu$ in (\ref{compute Y_D}) is replaced by $\mu_S$. In each iteration, $\mu_S$, $\mu_R$ and ${\bm \alpha}$ are updated as in (\ref{sub_gradient_typeI individual}) using the power allocation computed in (\ref{optimal power_typeII}) ($p_{k,m,1}^\ast$ computed by the RHS of (\ref{optimal power_typeI_individual}) multiplied by $s_{k,m}$). If there is an intermediate mode SP in the optimal solution, it must belong to the situation where the conditions (\ref{mode_conditions_individual}) are applicable. Then, like in Section~\ref{sec_dual_type_I_individual}, the zero-crossing method in \cite[Section 3.2]{relay_DF_individual_power_constraint} can be used to obtain the optimal $\mu_R/\mu_S$. The corresponding mode classification and power allocation can then be done accordingly.

\section{Simulation Results}\label{Sec_simulation}
This section provides the simulation results for the rates obtained by the proposed algorithms, and the dual optimum values which serve as the performance upper bounds.
The performances with fixed subcarrier pairing and with the SCP proposed in \cite{relay_eq_channel_gain} are also presented for comparison. Note that the original SCP in \cite{relay_eq_channel_gain} considered only the unweighted sum rate. It first sorts the subcarriers of the SR link and the RD link, respectively, according to their normalized channel gains, then pairs the SR subcarrier with the RD subcarrier having the same rank. For weighted sum rate, according to (\ref{impact_weighting}) and the discussion right after it, we modify the SCP such that $w_{k}a^{SR}_{k}$ and $a^{RD}_{m}$ are sorted first. Then the SR subcarrier and the RD subcarrier with the same rank are paired.
The RD link channel gains are not weighted for the reason that we do not know the actual subcarrier pairing scheme in advance.

\begin{figure}[htp]
\centering
\includegraphics[width=0.77\textwidth]{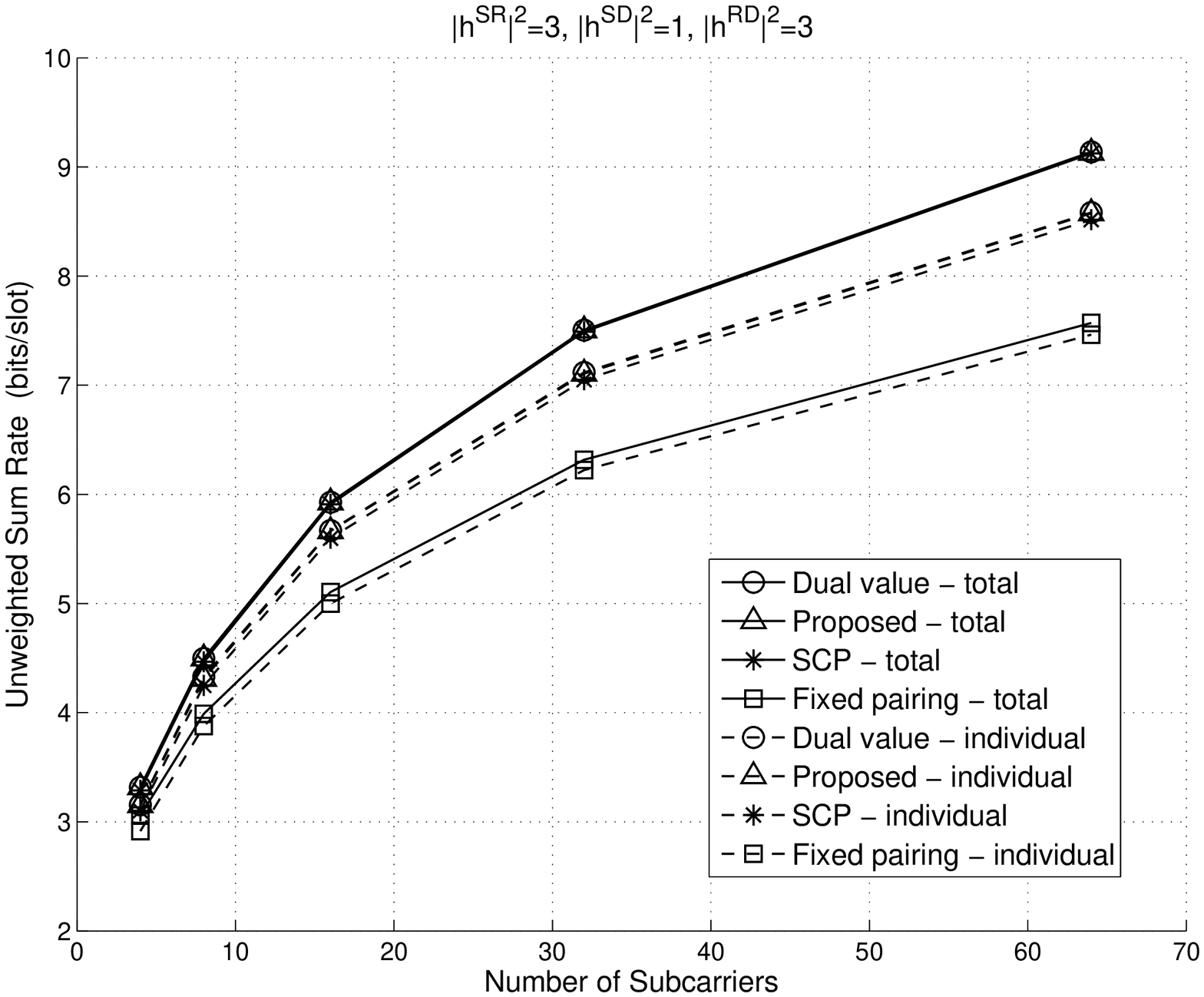}
\caption{Unweighted sum rates for the systems with $\mathbb{E}[|h^{SR}|^2]=3$, $\mathbb{E}[|h^{SD}|^2]=1$ and $\mathbb{E}[|h^{RD}|^2]=3$.}
\label{Fig_WSR_channel_3_1_3_noweight_total_individual}
\end{figure}

\begin{figure}[htp]
\centering
\includegraphics[width=0.77\textwidth]{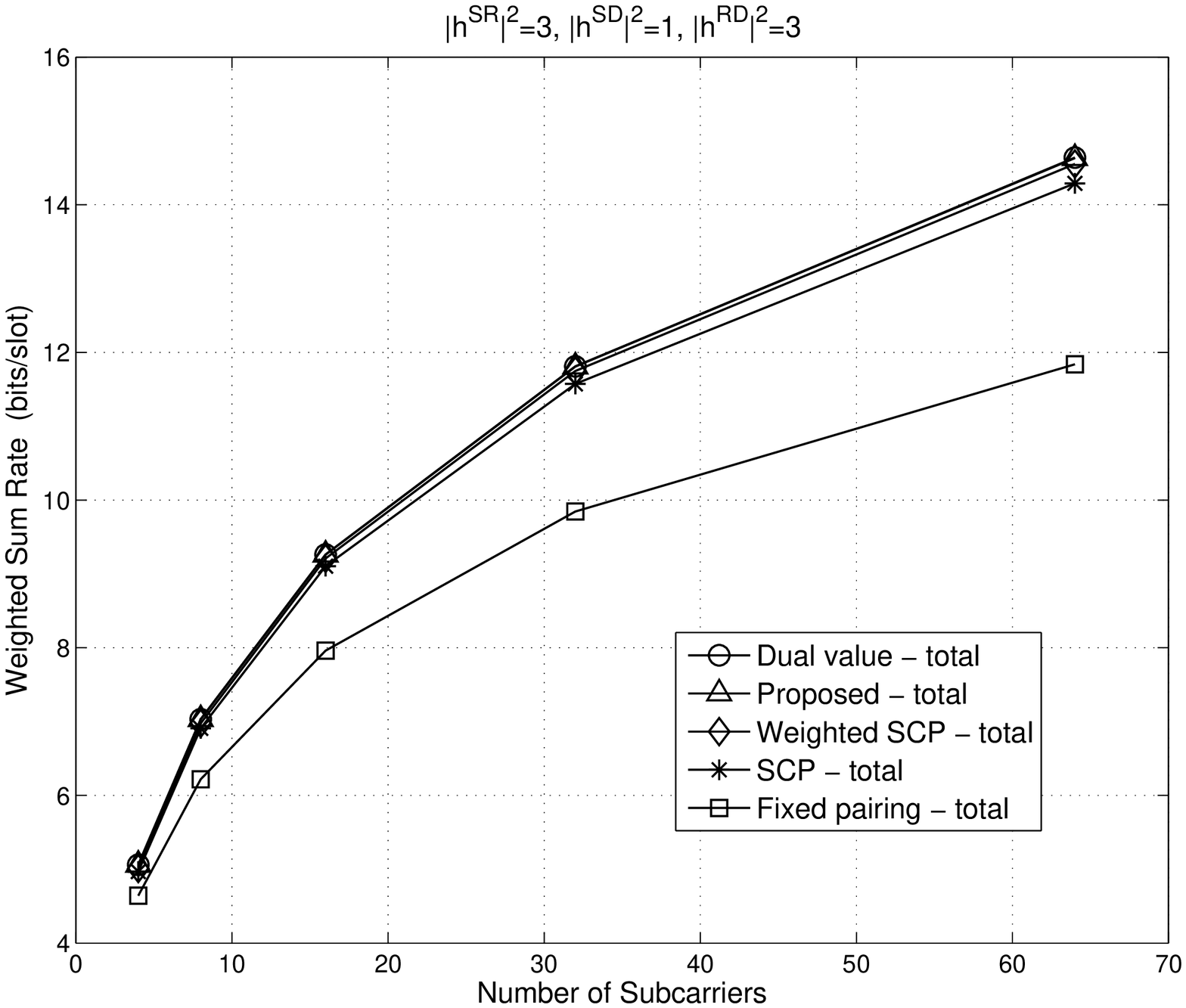}
\caption{Weighted sum rates for the systems with total power constraint, and $\mathbb{E}[|h^{SR}|^2]=3$, $\mathbb{E}[|h^{SD}|^2]=1$, $\mathbb{E}[|h^{RD}|^2]=3$.}
\label{Fig_WSR_channel_3_1_3_weight}
\end{figure}

\begin{figure}[htp]
\centering
\includegraphics[width=0.77\textwidth]{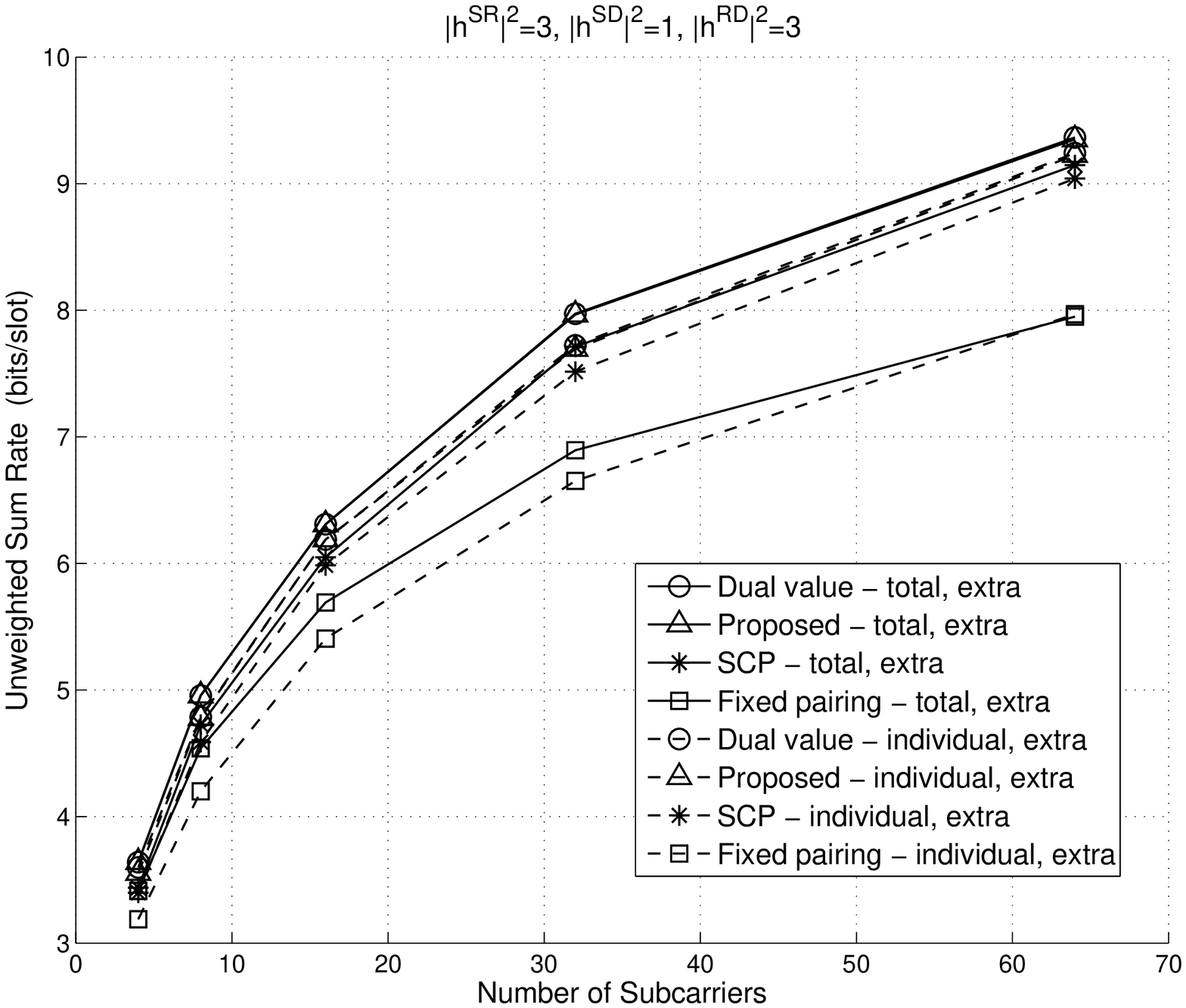}
\caption{Unweighted sum rates for the systems with extra direct-link transmission, and $\mathbb{E}[|h^{SR}|^2]=3$, $\mathbb{E}[|h^{SD}|^2]=1$, $\mathbb{E}[|h^{RD}|^2]=3$.}
\label{Fig_WSR_channel_3_1_3_noweight_total_individual_extra}
\end{figure}

\begin{figure}[htp]
\centering
\includegraphics[width=0.77\textwidth]{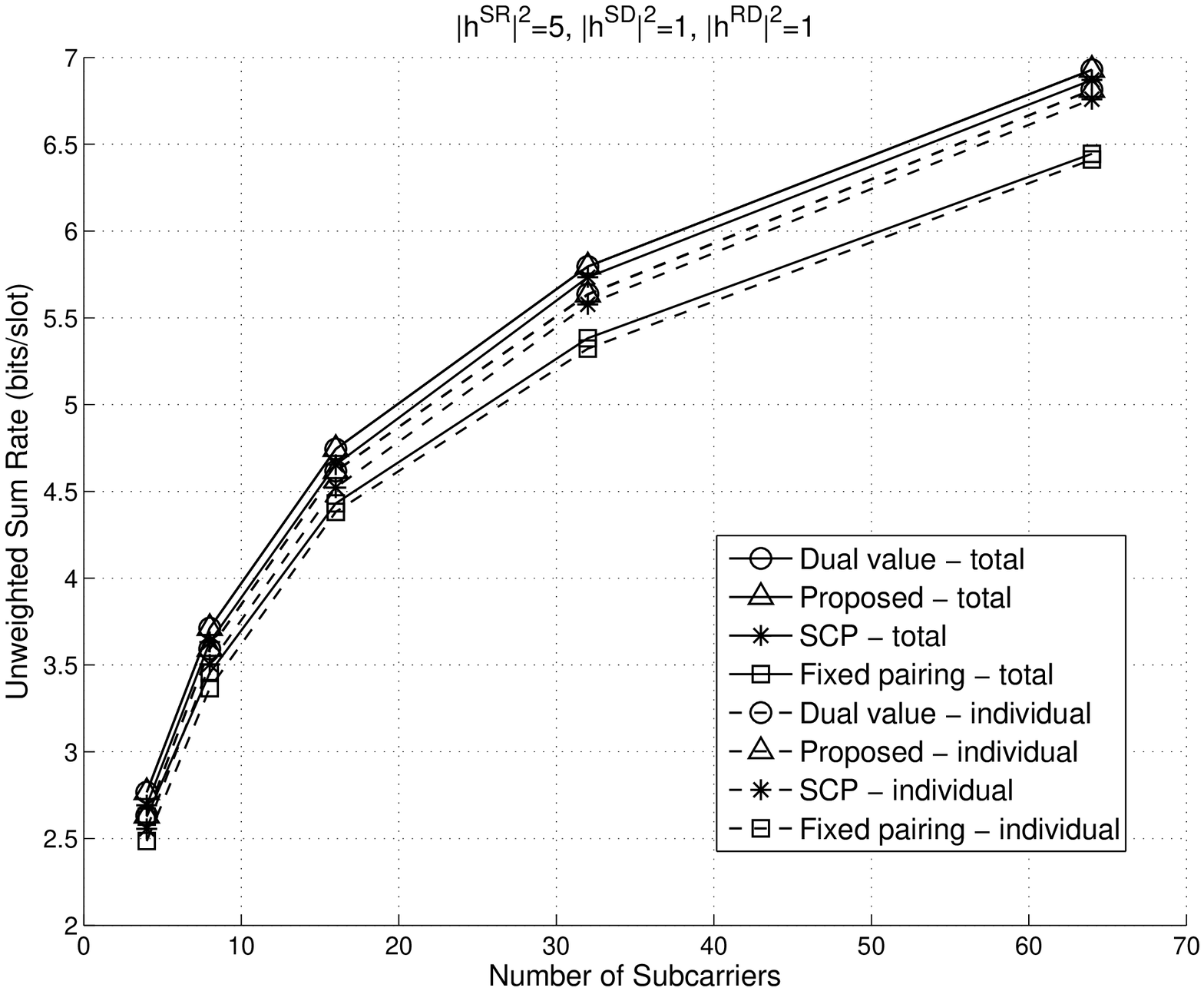}
\caption{Unweighted sum rates for the systems with $\mathbb{E}[|h^{SR}|^2]=5$, $\mathbb{E}[|h^{SD}|^2]=1$ and $\mathbb{E}[|h^{RD}|^2]=1$.}
\label{Fig_WSR_channel_5_1_1_noweight_total_individual}
\end{figure}

\begin{figure}[htp]
\centering
\includegraphics[width=0.77\textwidth]{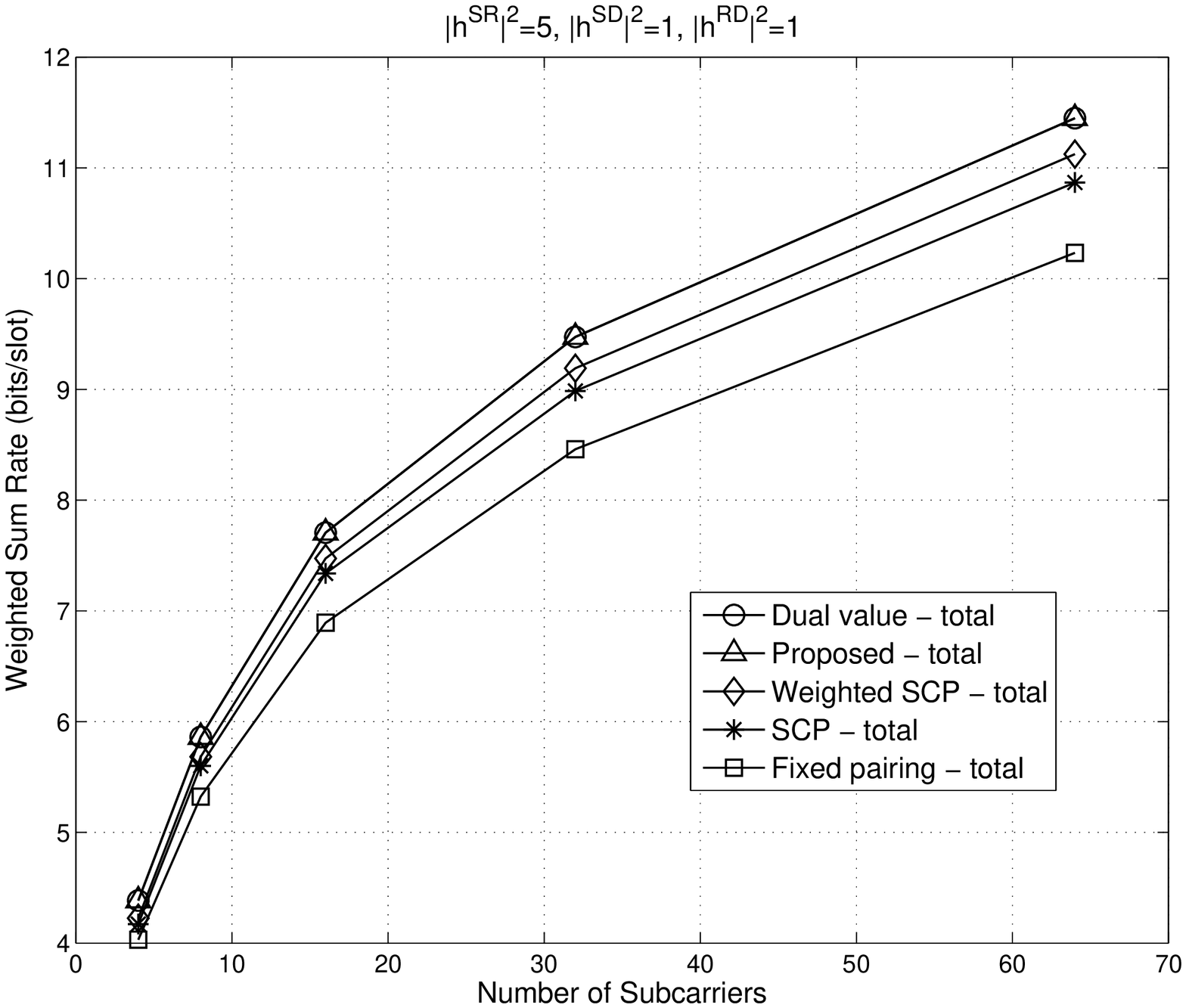}
\caption{Weighted sum rates for the systems with total power constraint, and $\mathbb{E}[|h^{SR}|^2]=5$, $\mathbb{E}[|h^{SD}|^2]=1$, $\mathbb{E}[|h^{RD}|^2]=1$.}
\label{Fig_WSR_channel_5_1_1_weight}
\end{figure}

\begin{figure}[htp]
\centering
\includegraphics[width=0.77\textwidth]{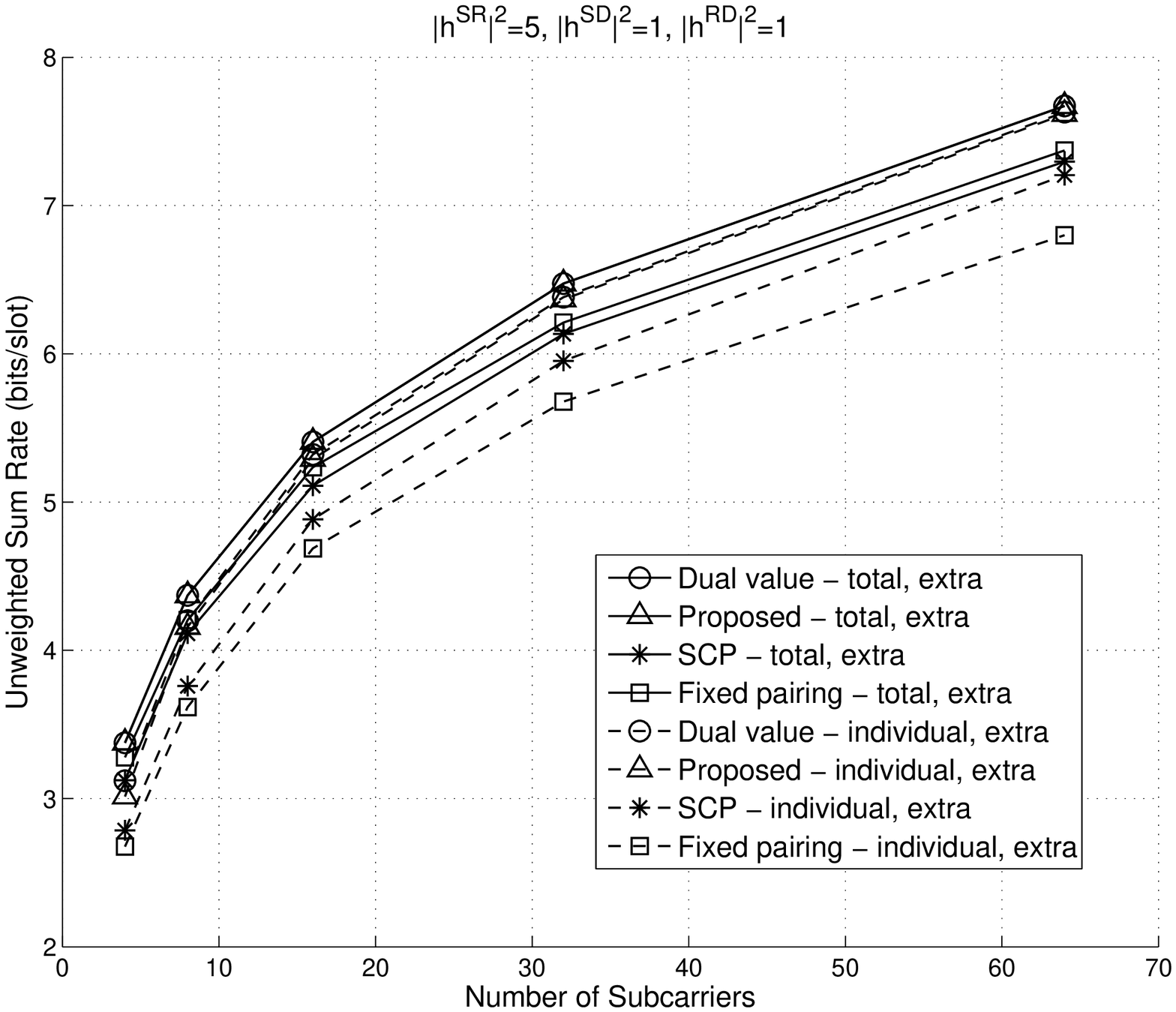}
\caption{Unweighted sum rates for the systems with extra direct-link transmission, and $\mathbb{E}[|h^{SR}|^2]=5$, $\mathbb{E}[|h^{SD}|^2]=1$, $\mathbb{E}[|h^{RD}|^2]=1$.}
\label{Fig_WSR_channel_5_1_1_noweight_total_individual_extra}
\end{figure}

\begin{figure}[htp]
\centering
\includegraphics[width=0.77\textwidth]{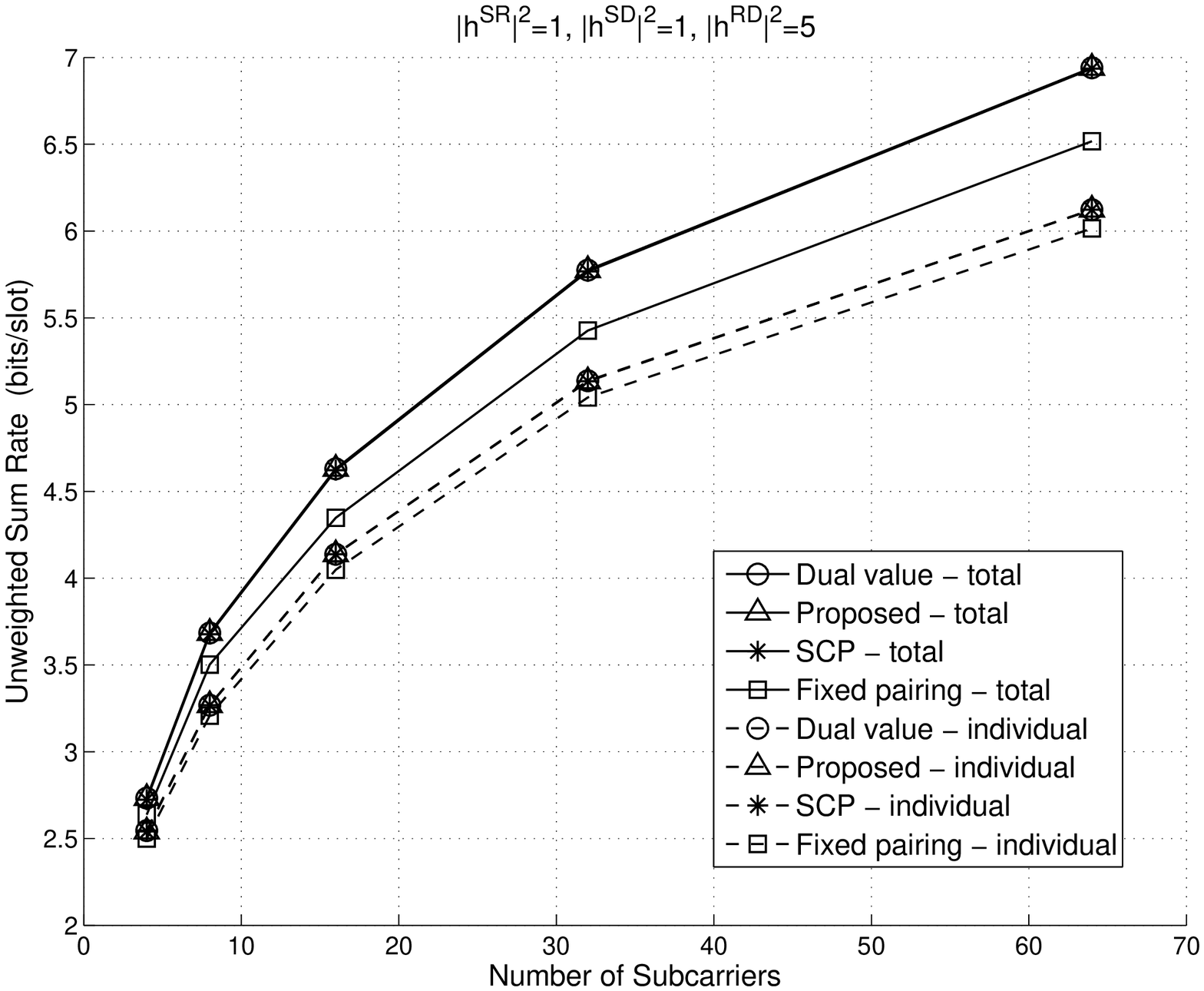}
\caption{Unweighted sum rates for the systems with $\mathbb{E}[|h^{SR}|^2]=1$, $\mathbb{E}[|h^{SD}|^2]=1$ and $\mathbb{E}[|h^{RD}|^2]=5$.}
\label{Fig_WSR_channel_1_1_5_noweight_total_individual}
\end{figure}

\begin{figure}[htp]
\centering
\includegraphics[width=0.77\textwidth]{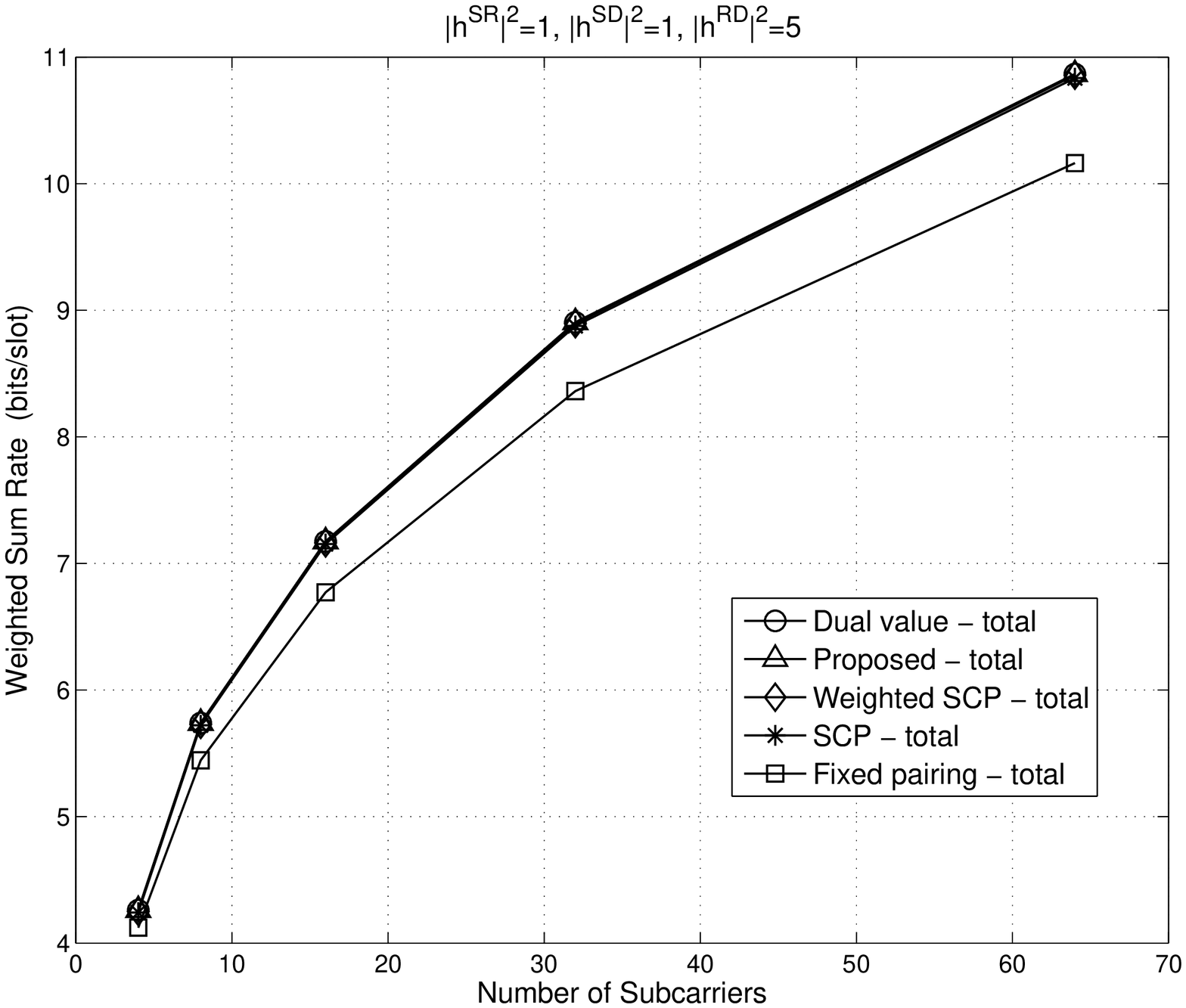}
\caption{Weighted sum rates for the systems with total power constraint, and $\mathbb{E}[|h^{SR}|^2]=1$, $\mathbb{E}[|h^{SD}|^2]=1$, $\mathbb{E}[|h^{RD}|^2]=5$.}
\label{Fig_WSR_channel_1_1_5_weight}
\end{figure}

\begin{figure}[htp]
\centering
\includegraphics[width=0.77\textwidth]{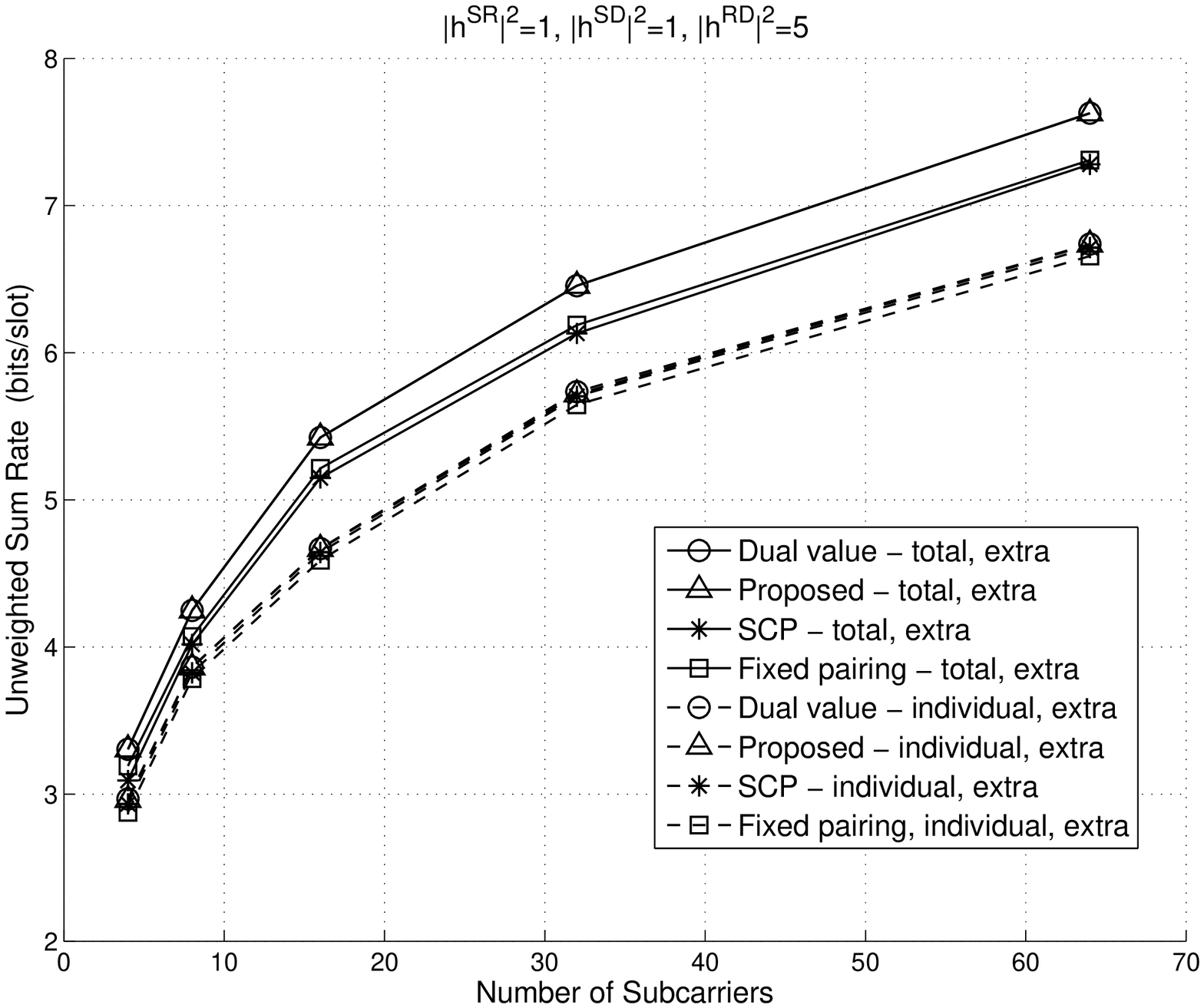}
\caption{Unweighted sum rates for the systems with extra direct-link transmission, and $\mathbb{E}[|h^{SR}|^2]=1$, $\mathbb{E}[|h^{SD}|^2]=1$, $\mathbb{E}[|h^{RD}|^2]=5$.}
\label{Fig_WSR_channel_1_1_5_noweight_total_individual_extra}
\end{figure}

The channels of different links are assumed to be independent of one another. The channels of the subcarriers are
independent and identically distributed (i.i.d.) Rician fading channels with $K$-factor $=1$. They are assumed constant within each two-slot period, and varying independently from one period to another. The AWGN variance is assumed to be one.
The total power constraint is set as $P=5$. As for the cases with individual power constraints, the
source power constraint is $P_S=4$ and the relay power constraint is $P_R=1$. These constraints are set with the practical consideration that the relay usually plays the role of assisting the transmission and/or extending the coverage, and has a smaller power than the source. In addition, when the relay is allowed more power and the achievable rate becomes limited by the source power constraint, some of the relay power will not be used. Setting $P_S=4$ and $P_R=1$ reduces the occurrence of this situation and makes the comparison with the total power constrained case fairer.
For all cases, the SD link is present, and the destination performs MRC whenever the relay is used. The SCP schemes first establish subcarrier pairing using SCP. Then, in the total power constrained cases (including the case with extra direct-link transmission), (\ref{relay_condition_total}) is used as the condition to use relay.
In the individual power constrained case without extra direct-link transmission, the method in \cite{relay_DF_individual_power_constraint} is used for mode classification and power allocation. For the individual power constrained case with extra direct-link transmission, the method in \cite{relay_DF_individual_power_constraint} cannot be used because the optimal mode classification conditions are no longer (\ref{mode_conditions_individual}). Naively using (\ref{mode_conditions_individual}) and (\ref{Louveaux_power_allocation}) may result in invalid power allocation as they are not the solutions in this case, and will affect the $\mu_S$, $\mu_R$ values through the iterations. On the other hand, modifying the method in \cite{Vandendorpe_J} to accommodate weighted rates is tedious. Thus, the algorithm discussed in Section~\ref{sec_extraSD_individual} is used with fixed subcarrier pairing from the SCP. The fixed pairing schemes use the same mode classification and power allocation procedures as that of the SCP schemes.
In Figs. \ref{Fig_WSR_channel_3_1_3_noweight_total_individual}, \ref{Fig_WSR_channel_3_1_3_noweight_total_individual_extra}, \ref{Fig_WSR_channel_5_1_1_noweight_total_individual}, \ref{Fig_WSR_channel_5_1_1_noweight_total_individual_extra}, \ref{Fig_WSR_channel_1_1_5_noweight_total_individual} and \ref{Fig_WSR_channel_1_1_5_noweight_total_individual_extra}, the unweighted sum rate is considered. That is, the all-one weighting factor is used.
In Figs. \ref{Fig_WSR_channel_3_1_3_weight}, \ref{Fig_WSR_channel_5_1_1_weight} and \ref{Fig_WSR_channel_1_1_5_weight}, the weighted sum rate is considered with $w_{k}=1+\frac{k-1}{M-1},\forall k$, which is used only as an example with a concise expression.

For the proposed algorithms, $\mu$, $\mu_R$, $\mu_S$ and $\alpha_m$'s were randomly initialized to be between 0 and 2 for each two-slot period. For each number of subcarriers $(\in \{4, 8, 16, 32, 64\})$, 1000 such two-slot periods were simulated, and the results averaged to avoid favoring certain initial conditions. The step sizes for the subgradient method were all set as $\frac{0.05}{\sqrt{i}}$, where $i$ is the iteration index. In the simulation, we observed that the number of iterations before the amendment algorithm was triggered depends on the number of subcarriers. The number of iterations needed ranged roughly from a few hundreds for small numbers of subcarriers ($<10$) to slightly more than 10000 for 64 subcarriers.

We investigate three system configurations corresponding to different scenarios. In Figs. \ref{Fig_WSR_channel_3_1_3_noweight_total_individual}, \ref{Fig_WSR_channel_3_1_3_weight} and \ref{Fig_WSR_channel_3_1_3_noweight_total_individual_extra}, the mean square channel gains of the SR, SD and RD links are 3, 1, 3, respectively. This corresponds to the situation where the relay is placed between the source and the destination. In Figs. \ref{Fig_WSR_channel_5_1_1_noweight_total_individual}, \ref{Fig_WSR_channel_5_1_1_weight} and \ref{Fig_WSR_channel_5_1_1_noweight_total_individual_extra}, the mean square channel gains of the SR, SD and RD links are 5, 1 and 1, respectively, which means that the relay is close to the source. In Figs. \ref{Fig_WSR_channel_1_1_5_noweight_total_individual}, \ref{Fig_WSR_channel_1_1_5_weight} and \ref{Fig_WSR_channel_1_1_5_noweight_total_individual_extra}, the mean square channel gains of the SR, SD and RD links are 1, 1 and 5, respectively, which means that the relay is close to the destination.
In these figures, we find that, in all cases, the rates obtained by the proposed algorithms are almost equal to the corresponding dual optimum values. This validates the arguments in Section~\ref{duality_gap_total}, Section~\ref{sec_dual_type_I_individual} and Section~\ref{sec_extraSD_total} that the duality gap is virtually zero when the number of subcarriers is reasonably large. Even when the number of subcarriers is 4, the duality gap is hardly noticeable from the averaged results, because it is zero with a very high probability.
These results also show that the proposed algorithms can almost achieve the optimal weighted sum rates. There are some other general trends that can be observed from these figures. One of them is that fixed subcarrier pairing incurs a significant performance loss. In addition, the weighted and unweighted sum rates increase with the number of subcarriers due to frequency diversity and more flexibility in pairing. As to the performance under different constraints, the performance under total power constraint is better than the performance under individual power constraints, due to the flexibility in power allocation. By comparing Figs. \ref{Fig_WSR_channel_3_1_3_noweight_total_individual} and \ref{Fig_WSR_channel_3_1_3_noweight_total_individual_extra}, \ref{Fig_WSR_channel_5_1_1_noweight_total_individual} and \ref{Fig_WSR_channel_5_1_1_noweight_total_individual_extra}, \ref{Fig_WSR_channel_1_1_5_noweight_total_individual} and \ref{Fig_WSR_channel_1_1_5_noweight_total_individual_extra}, it is clear that extra direct-link transmission always improves the performance.

The SCP was proved in \cite{relay_AF_DF_eq_channel_gain}\cite{Li_AFDF_OFDM} to be optimal for the unweighted system without the SD link
under the total power constraint. When the SD link is present and/or when weighted sum rate is considered, the performance of the SCP depends on the link qualities. The SCP almost achieves the optimal unweighted sum rate
for the cases with total power constraint and no extra direct-link transmission in Fig.~\ref{Fig_WSR_channel_3_1_3_noweight_total_individual} and Fig.~\ref{Fig_WSR_channel_1_1_5_noweight_total_individual}, but becomes noticeably worse than the optimal in Fig.~\ref{Fig_WSR_channel_5_1_1_noweight_total_individual}. For the scenario in Fig.~\ref{Fig_WSR_channel_3_1_3_noweight_total_individual}, this is reasonable because the SD link is relatively weak compared to the other two links. Thus the direct-link mode is rarely used, and the SCP is nearly optimal for the relay-mode SPs given that their SD subcarriers are weak. For the scenario in Fig.~\ref{Fig_WSR_channel_1_1_5_noweight_total_individual}, the RD link is the strongest and seldom becomes the bottleneck for mode selection. For the SCP as well as the proposed algorithm, mode selection is mainly determined by the SR and SD links. For the direct-link mode SPs, the SCP and the proposed algorithm have similar performances. For the relay mode SPs, the SCP is nearly optimal because the SD link is the weakest among the three links. Overall, the SCP has a very similar performance to that of the proposed algorithm which is almost optimal. As to the case of Fig.~\ref{Fig_WSR_channel_5_1_1_noweight_total_individual}, we can see that since the SR link is much stronger than the SD link, the condition for using relay (\ref{relay_condition_total}) is dominated by the relation between the channel gains of the SD and the RD links. However, the SCP does not consider the SD link in establishing subcarrier pairing. As a result, the SCP is almost equivalent to random pairing in terms of optimizing the mode selection and sum rate. Thus its sum rate is smaller than that of the proposed algorithm.
The SCP is still better than fixed pairing because it helps the SPs that are in the relay mode.

For the individual power constrained cases, or when weighted sum rate is considered, as shown in Figs.~\ref{Fig_WSR_channel_3_1_3_noweight_total_individual}, \ref{Fig_WSR_channel_5_1_1_noweight_total_individual}, and \ref{Fig_WSR_channel_3_1_3_weight}, \ref{Fig_WSR_channel_5_1_1_weight}, the gaps between the SCP and the proposed algorithms become larger. This is due to the mismatches between the SCP and these scenarios. To show that our modification to the original SCP is meaningful, we show the performance of the original (unweighted) SCP together with that of the ``weighted SCP" in Fig.~\ref{Fig_WSR_channel_3_1_3_weight} and Fig.~\ref{Fig_WSR_channel_5_1_1_weight}. These two figures clearly show that the original SCP is not suitable when weighted sum rate is considered. The performance gap between the ``weighted SCP" and the proposed algorithm in Fig.~\ref{Fig_WSR_channel_5_1_1_weight} is due to the aforementioned ``random pairing" effect of the SCP (as in the total power constrained case in Fig.~\ref{Fig_WSR_channel_5_1_1_noweight_total_individual}). However, these trends do not appear in Fig.~\ref{Fig_WSR_channel_1_1_5_noweight_total_individual} and Fig.~\ref{Fig_WSR_channel_1_1_5_weight}. For Fig.~\ref{Fig_WSR_channel_1_1_5_noweight_total_individual}, this is because the strong RD link makes the sum rate not limited by the low relay power constraint.
Therefore, for the SCP, the situation is very similar to that with the total power constraint. As a result, the SCP is almost optimal. For Fig.~\ref{Fig_WSR_channel_1_1_5_weight}, the strong RD link makes mode selection dependent almost only on the channel gains of the SR and SD links. Thus, mode selection is almost independent of the pairing scheme and weighting factors. For the source subcarriers that have relatively lower SR gains and are in the direct-link mode, all schemes yield similar performances. On the other hand, for the subcarriers in the relay mode, pairing better RD subcarriers with SR subcarriers having higher weighted channel gains can improve the weighted sum rate. Both the original SCP and the weighted SCP can do that for the SR subcarriers that are strong enough. Thus they both perform well and almost optimally.

With possible extra direct-link transmission, the SCP is worse than the proposed algorithm for not considering the benefits of the extra direct-link transmission (such as more diversity from the additional independent channels, and more flexibility in water-filling) in subcarrier pairing and mode selection.
Under the total power constraint, we find that the SCP is similar and even slightly worse than fixed pairing in Fig.~\ref{Fig_WSR_channel_5_1_1_noweight_total_individual_extra} and Fig. \ref{Fig_WSR_channel_1_1_5_noweight_total_individual_extra}. This is because, without considering the possible extra direct-link transmission, the SCP's pairing of strong SR subcarrier with strong RD subcarrier tends to satisfy (\ref{relay_condition_total}) more than fixed pairing, and make more SPs use the relay. Thus it loses the opportunities to transmit more messages with the extra direct-link. This phenomenon does not appear in Fig.~\ref{Fig_WSR_channel_3_1_3_noweight_total_individual_extra}, for which the benefits of the extra direct-link transmission are not significant due to the weak SD link. Under individual power constraints, both the SCP and the fixed pairing schemes use the algorithm in Section \ref{sec_extraSD_individual} for optimal joint mode selection and power allocation. The SCP always performs better than fixed pairing due to its better subcarrier pairing. In Fig.~\ref{Fig_WSR_channel_1_1_5_noweight_total_individual_extra}, the advantage of the SCP over fixed pairing is smaller than in Fig.~\ref{Fig_WSR_channel_1_1_5_noweight_total_individual} because the optimal mode selection assigns more SPs to the direct-link mode for which better SR-RD subcarrier pairing does not improve the rate.

\section{Conclusion}\label{Sec_conclusion}
In this paper we investigated OFDM point to point transmission, enhanced with a DF relay. We jointly optimized
subcarrier pairing and power allocation to maximize the weighted sum rate with consideration of the source-destination link and destination combining. To the best of our knowledge, this problem has not been solved before.
Both total power constraint and individual power constraints for the source and the relay were considered.
The system that allows additional messages to be transmitted on the idle subcarriers not used by the relay, in the source-destination link in the second time slot, was also investigated.
We solved the optimization problems by using some special properties of the systems, as well the continuous relaxation and the dual method.
The subgradient method was adopted to find the Lagrange multipliers which also
helped us to find the primal feasible solutions.
Based on the optimization results, algorithms with tractable complexities to obtain feasible subcarrier pairing schemes and the corresponding
power allocations were proposed.
Simulation results showed that the proposed algorithms can achieve nearly optimal weighted sum rates, and outperform the method proposed in \cite{relay_eq_channel_gain} under various channel conditions.

\bibliographystyle{IEEEtran}
\bibliography{IEEEabrv,aning_thesis}

\begin{thebibliography}{10}
\providecommand{\url}[1]{#1}
\csname url@samestyle\endcsname
\providecommand{\newblock}{\relax}
\providecommand{\bibinfo}[2]{#2}
\providecommand{\BIBentrySTDinterwordspacing}{\spaceskip=0pt\relax}
\providecommand{\BIBentryALTinterwordstretchfactor}{4}
\providecommand{\BIBentryALTinterwordspacing}{\spaceskip=\fontdimen2\font plus
\BIBentryALTinterwordstretchfactor\fontdimen3\font minus
  \fontdimen4\font\relax}
\providecommand{\BIBforeignlanguage}[2]{{%
\expandafter\ifx\csname l@#1\endcsname\relax
\typeout{** WARNING: IEEEtran.bst: No hyphenation pattern has been}%
\typeout{** loaded for the language `#1'. Using the pattern for}%
\typeout{** the default language instead.}%
\else
\language=\csname l@#1\endcsname
\fi
#2}}
\providecommand{\BIBdecl}{\relax}
\BIBdecl

\bibitem{DF_individual_downlink}
W.~Nam, W.~Chang, S.~Y. Chung, and Y.~H. Lee, ``{Transmit Optimization for
  Relay-based Cellular OFDMA Systems},'' in \emph{Proc. IEEE Int. Conf. on
  Commun. (ICC'07)}, Jun. 2007, pp. 5714--5719.

\bibitem{DF_OFDM_total_individual_power}
L.~Vandendorpe, R.~Duran, J.~Louveaux, and A.~Zaidi, ``{Power Allocation for
  OFDM Transmission with DF Relaying},'' in \emph{Proc. IEEE Int. Conf. on
  Commun. (ICC'08)}, May 2008, pp. 3795--3800.

\bibitem{relay_DF_individual_power_constraint}
J.~Louveaux, R.~Duran, and L.~Vandendorpe, ``{Efficient Algorithm for Optimal
  Power Allocation in OFDM Transmission with Relaying},'' in \emph{Proc. IEEE
  Int. Conf. on Acoustics, Speech, and Signal Processing}, Mar. 2008, pp.
  3257--3260.

\bibitem{Vandendorpe_J}
L.~Vandendorpe, J.~Louveaux, O.~Oguz, and A.~Zaidi, ``{Rate-Optimized Power
  Allocation for {DF}-Relayed {OFDM} Transmission under Sum and Individual
  Power Constraints},'' \emph{EURASIP Journal on Wireless Communications and
  Networking}, vol. 2009, Article ID 814278, 2009.

\bibitem{relay_eq_channel_gain}
Y.~Wang, X.~Qu, T.~Wu, and B.~Liu, ``{Power Allocation and Subcarrier Pairing
  Algorithm for Regenerative OFDM Relay System},'' in \emph{Proc. IEEE Vehic.
  Technol. Conf. (VTC'07)}, Apr. 2007, pp. 2727--2731.

\bibitem{relay_AF_DF_eq_channel_gain}
Y.~Li, W.~Wang, J.~Kong, W.~Hong, X.~Zhang, and M.~Peng, ``{Power Allocation
  and Subcarrier Pairing in OFDM-Based Relaying Networks},'' in \emph{Proc.
  IEEE Int. Conf. on Commun. (ICC'08)}, May 2008, pp. 2602--2606.

\bibitem{Li_AFDF_OFDM}
Y.~Li, W.~Wang, J.~Kong, and M.~Peng, ``{Subcarrier Pairing for
  Amplify-and-Forward and Decode-and-Forward {OFDM} Relay Links},''
  \emph{{IEEE} Commun. Lett.}, vol.~13, no.~4, pp. 209--211, Apr. 2009.

\bibitem{Herdin}
M.~Herdin, ``{A Chunk Based OFDM Amplify-and-Forward Relaying Scheme for 4G
  Mobile Radio Systems},'' in \emph{Proc. IEEE Int. Conf. on Commun. (ICC'06)},
  Jun. 2006, pp. 4507--4512.

\bibitem{wittneben}
I.~Hammerstrom and A.~Wittneben, ``{Joint Power Allocation for Nonregenerative
  MIMO-OFDM Relay Links},'' in \emph{Proc. IEEE Int. Conf. Acoustic, Speech,
  and Signal Processing (ICASSP'06)}, May 2006, pp. 49--52.

\bibitem{AF_pairing_without_diversity}
------, ``{Power Allocation Schemes for Amplify-and-Forward MIMO-OFDM Relay
  Links},'' \emph{{IEEE} Trans. Wireless Commun.}, vol.~6, no.~8, pp.
  2798--2802, Aug. 2007.

\bibitem{relay_AF_OPT_SUBCHANNEL_ASSIGNMENT}
A.~Hottinen and T.~Heikkinen, ``{Optimal Subchannel Assignment in a Two-hop
  OFDM Relay},'' in \emph{IEEE Int. Workshop on Signal Processing Adv. Wireless
  Commun. (SPAWC'07)}, 2007, pp. 104--111.

\bibitem{AF_OFDM_total_individual_power}
I.~Hammerstrom and A.~Wittneben, ``{On the Optimal Power Allocation for
  Nonregenerative OFDM Relay Links},'' in \emph{Proc. IEEE Int. Conf. on
  Commun. (ICC'06)}, Jun. 2006, pp. 4463--4468.

\bibitem{relay_AF_total_optimal_power}
M.~Saito, C.~Athaudage, and J.~Evans, ``{On Power Allocation for Dual-Hop
  Amplify-and-Forward OFDM Relay Systems},'' in \emph{Proc. IEEE Global Commun.
  Conf. (GLOBECOM)}, Nov. 2008, pp. 4419 -- 4423.

\bibitem{concave_function}
W.~Yu and J.~M. Cioffi, ``{FDMA Capacity of Gaussian Multiple-Access Channels
  with ISI},'' \emph{{IEEE} Trans. Commun.}, vol.~50, no.~1, pp. 102--111, Aug.
  2002.

\bibitem{subgradient_method}
S.~Boyd and A.~Mutapcic, ``{Subgradient Methods},'' in \emph{lecture notes of
  EE364b, {S}tanford Univ. Spring Quarter 2007 - 08}.

\bibitem{Yu_dual}
W.~Yu and R.~Lui, ``{Dual Methods for Nonconvex Spectrum Optimization of Multi-
  carrier Systems},'' \emph{{IEEE} Trans. Commun.}, vol.~54, no.~7, pp.
  1310--1322, Jul. 2006.

\bibitem{zero_gap_SUB_infinity}
K.~Seong, M.~Mohseni, and J.~Cioffi, ``{Optimal Resource Allocation for OFDMA
  Downlink Systems},'' in \emph{Proc. IEEE Int. Symp. Inform. Theory
  (ISIT'06)}, Jul. 2006, pp. 1394--1398.

\bibitem{relay_capacity_theorem}
T.~Cover and A.~Gamal, ``{Capacity Theorems for the Relay Channel},''
  \emph{{IEEE} Trans. Inf. Theory}, vol.~25, no.~5, pp. 572--584, Jan. 1979.

\bibitem{multilevel_waterfilling}
L.~M.~C. Hoo, B.~Halder, J.~Tellado, and J.~M. Cioffi, ``{Multiuser Transmit
  Optimization for Multicarrier Broadcast Channels: Asymptotic FDMA Capacity
  Region and Algorithms},'' \emph{{IEEE} Trans. Commun.}, vol.~52, no.~6, pp.
  922--930, Jun. 2004.

\bibitem{Book_Boyd}
S.~Boyd and L.~Vandenberghe, \emph{{Convex Optimization}}.\hskip 1em plus 0.5em
  minus 0.4em\relax Cambridge University Press, 2004.

\bibitem{Chiang_zero_gap}
P.~Hande, S.~Zhang, and M.~Chiang, ``{Distributed Rate Allocation for Inelastic
  Flows},'' \emph{{IEEE/ACM} Trans. Netw.}, vol.~15, no.~6, pp. 1240--1253,
  Dec. 2007.

\end{thebibliography}


\begin{table}
\renewcommand{\arraystretch}{1.3}
\caption{Algorithm for the total power constrained case.}
\label{Table_algo_obtain_feasible_typeI} \centering
\begin{tabular}{lll}
\multicolumn{3}{l}{\textit{Initialize }$i=1, ~\mu^{(1)}, ~{\bm \alpha}^{(1)}, ~\epsilon=0.01$, amendment = \textit{false}, max\_it = $\infty$, sum\_rate = 0}\\
 \multicolumn{3}{l}{\textit{While }($i<$ max\_it) ~~~~~~~\textit{(** iteration i **)}}\\
 &\textit{Compute }$X_{k,m}^{(i)}, \forall k, m$, \textit{ using }$\mu^{(i)}$, ${\bm \alpha}^{(i)}$ \textit{ in } (\ref{compute X})\\
 &$\textit{Compute }{\bm t}^{(i)}\textit{ using }X_{k,m}^{(i)} \textit{ in } (\ref{optimal_t_typeI})$\\
 &$\textit{Compute }{\bm p}^{(i)}\textit{ using }\mu^{(i)}, ~{\bm t}^{(i)} \textit{ in }(\ref{optimal power_typeI})$\\
 &$\textit{Compute }\mu^{(i+1)}, ~{\bm \alpha}^{(i+1)}\textit{ using }\mu^{(i)}, ~{\bm \alpha}^{(i)}, ~{\bm t}^{(i)}, ~{\bm p}^{(i)} \textit{ in }(\ref{sub_gradient_typeI_total})$\\
 &\textit{If} (amendment = \textit{false}) \textit{and} $\left(\frac{\lvert\mu^{(i+1)}-\mu^{(i)}\rvert}{\lvert \mu^{(i+1)} \rvert}<\epsilon\right)$ \textit{and} $\left(\frac{\lVert{\bm \alpha}^{(i+1)}-{\bm \alpha}^{(i)}\rVert}{\lVert {\bm \alpha}^{(i+1)} \rVert}<\epsilon\right)$\\
 &\hspace{5 mm}amendment = \textit{true}\\
 &\hspace{5 mm}max\_it = $\lfloor 1.1 \times i \rfloor$\\
 &\textit{End}\\
 &\textit{If} (amendment = \textit{true}) ~~~~~~~\textit{(** amendment algorithm **)}\\
 &\hspace{5 mm}$\hat{{\bm t}} = {\bm t}^{(i)}$, $c_m=\sum_{k=1}^M\hat{t}_{k,m}, \forall m$\\
 &\hspace{5 mm}\textit{For (j = 1 to M)}&\\
 &\hspace{7 mm}\textit{If } $(c_j>1)$, ~~$s^*=\arg\max_{\left\{s|\hat{t}_{s,j}=1\right\}}X_{s,j}^{(i)}$, ~~\textit{End}\\
 &\hspace{7 mm}\textit{While }($c_j>1$) \\
 &\hspace{10 mm}$m^*=\arg\min_{\{m|c_m=0\}}\left|\alpha^{(i)}_j-\alpha^{(i)}_m\right|$\\
 &\hspace{10 mm}$r^*=\arg\max_{\left\{r|\hat{t}_{r,j}=1, ~r\neq s^*\right\}}X_{r,m^*}^{(i)}$\\
 &\hspace{10 mm}$\hat{t}_{r^*,j}=0$, ~$\hat{t}_{r^*,m^*}=1$\\
 &\hspace{10 mm}$c_{j}=c_{j}-1, ~c_{m^*}=c_{m^*}+1$\\
 &\hspace{7 mm}\textit{End}\\
 &\hspace{5 mm}\textit{End}\\
 &\hspace{5 mm}\textit{With fixed subcarrier pairing} $\hat{{\bm t}},$\textit{ apply water-filling on the subcarrier pairs with equivalent}\\
 &\hspace{5 mm}~~\textit{channel gains in (\ref{Eq_ECG}) to compute power allocation }$\hat{{\bm p}}$\textit{ and the weighted sum rate} $R$\textit{ as in (\ref{objective_Rate_I_MIP})}.\\
 &\hspace{5 mm}\textit{If } ($R>$ sum\_rate), ~~sum\_rate $=R$, ~$\check{{\bm t}}=\hat{{\bm t}}$, ~$\check{{\bm p}}=\hat{{\bm p}}$, ~~\textit{End}\\
 &\textit{End ~~~~~~~(** amendment algorithm **)}\\
 &$i=i+1$\\
\multicolumn{3}{l}{\textit{End ~~~~~(** iteration i **)}}\\
\multicolumn{3}{l}{sum\_rate, $\check{{\bm t}}$ and $\check{{\bm p}}$ \textit{are the obtained weighted sum rate, feasible subcarrier pairing and power allocation, respectively.}}
\end{tabular}
\end{table}

\begin{table}
\renewcommand{\arraystretch}{1.3}
\caption{Algorithm for the individual power constrained case.}
\label{Table_algo_obtain_feasible_typeI_individual} \centering
\begin{tabular}{lll}
\multicolumn{3}{l}{\textit{Initialize }$i=1, ~\mu_S^{(1)}, ~\mu_R^{(1)}, ~{\bm \alpha}^{(1)}, ~\epsilon=0.01$, amendment = \textit{false}, max\_it = $\infty$, sum\_rate = 0}\\
 \multicolumn{3}{l}{\textit{While }($i<$ max\_it) ~~~~~~~\textit{(** iteration i **)}}\\
 &\textit{Determine modes for all possible subcarrier pairs, and obtain }$a_{k,m}$, $c^S_{k,m}$ and $c^R_{k,m}$\\
 &~~\textit{by using }$\mu_S^{(i)}$, $\mu_R^{(i)}$ \textit{ in } (\ref{Eq_ECG_individual}), (\ref{source_power_ratio_individual}) and (\ref{relay_power_ratio_individual})\\
 &\textit{Compute }$Z_{k,m}^{(i)}, \forall k, m$, \textit{ using }$\mu_S^{(i)}$, $\mu_R^{(i)}$, ${\bm \alpha}^{(i)}$, $a_{k,m}$, $c^S_{k,m}$, $c^R_{k,m}$ \textit{ in } (\ref{compute Z})\\
 &$\textit{Compute }{\bm t}^{(i)}\textit{ using }Z_{k,m}^{(i)} \textit{ in } (\ref{optimal_t_typeI_individual})$\\
 &$\textit{Compute }{\bm p}^{(i)}\textit{ using }\mu_S^{(i)}, ~\mu_R^{(i)}, ~{\bm t}^{(i)}, ~a_{k,m}, ~c^S_{k,m}, ~c^R_{k,m} \textit{ in }(\ref{optimal power_typeI_individual})$\\
 &$\textit{Compute }\mu_S^{(i+1)}, ~\mu_R^{(i+1)}, ~{\bm \alpha}^{(i+1)}\textit{ using }\mu_S^{(i)}, ~\mu_R^{(i)}, ~{\bm \alpha}^{(i)}, ~{\bm t}^{(i)}, ~{\bm p}^{(i)}, ~c^S_{k,m}, ~c^R_{k,m} \textit{ in }(\ref{sub_gradient_typeI individual})$\\
 &\textit{If} (amendment = \textit{false}) \textit{and} $\left(\frac{\lvert\mu_S^{(i+1)}-\mu_S^{(i)}\rvert}{\lvert \mu_S^{(i+1)} \rvert}<\epsilon\right)$ \textit{and} $\left(\frac{\lvert\mu_R^{(i+1)}-\mu_R^{(i)}\rvert}{\lvert \mu_R^{(i+1)} \rvert}<\epsilon\right)$ \textit{and} $\left(\frac{\lVert {\bm \alpha}^{(i+1)}-{\bm \alpha}^{(i)}\rVert}{\lVert {\bm \alpha}^{(i+1)} \rVert}<\epsilon\right)$\\
 &\hspace{5 mm}amendment = \textit{true}\\
 &\hspace{5 mm}max\_it = $\lfloor 1.1 \times i \rfloor$\\
 &\textit{End}\\
 &\textit{If} (amendment = \textit{true}) ~~~~~~~\textit{(** amendment algorithm **)}\\
 &\hspace{5 mm}$\hat{{\bm t}} = {\bm t}^{(i)}$, $c_m=\sum_{k=1}^M\hat{t}_{k,m}, \forall m$\\
 &\hspace{5 mm}\textit{For (j = 1 to M)}&\\
 &\hspace{7 mm}\textit{If } $(c_j>1)$, ~~$s^*=\arg\max_{\left\{s|\hat{t}_{s,j}=1\right\}}Z_{s,j}^{(i)}$, ~~\textit{End}\\
 &\hspace{7 mm}\textit{While }($c_j>1$) \\
 &\hspace{10 mm}$m^*=\arg\min_{\{m|c_m=0\}}\left|\alpha^{(i)}_j-\alpha^{(i)}_m\right|$\\
 &\hspace{10 mm}$r^*=\arg\max_{\left\{r|\hat{t}_{r,j}=1, ~r\neq s^*\right\}}Z_{r,m^*}^{(i)}$\\
 &\hspace{10 mm}$\hat{t}_{r^*,j}=0$, ~$\hat{t}_{r^*,m^*}=1$\\
 &\hspace{10 mm}$c_{j}=c_{j}-1, ~c_{m^*}=c_{m^*}+1$\\
 &\hspace{7 mm}\textit{End}\\
 &\hspace{5 mm}\textit{End}\\
 &\hspace{5 mm}\textit{With fixed subcarrier pairing} $\hat{{\bm t}},$\textit{ and letting }$\hat{\mu}_S=\mu_S^{(i)}$, $\hat{\mu}_R=\mu_R^{(i)}$, \textit{ use the zero-crossing method in \cite[Section 3.2]{relay_DF_individual_power_constraint}}\\
 &\hspace{5 mm}~~\textit{to update $\hat{\mu}_S$, $\hat{\mu}_R$ to their optimal.}\\
 &\hspace{5 mm}\textit{Use $\hat{\mu}_S$, $\hat{\mu}_R$ in (\ref{mode_conditions_individual}) and (\ref{Louveaux_power_allocation}) to obtain mode classification and power allocation} $\hat{{\bm p}}_{S}$, $\hat{{\bm p}}_{R}$,\\
 &\hspace{5 mm}~~\textit{and compute the weighted sum rate $R$}.\\
 &\hspace{5 mm}\textit{If } ($R>$ sum\_rate), ~~sum\_rate $=R$, ~$\check{{\bm t}}=\hat{{\bm t}}$, ~$\check{{\bm p}}_{S}=\hat{{\bm p}}_{S}$, ~$\check{{\bm p}}_{R}=\hat{{\bm p}}_{R}$, ~~\textit{End}\\
 &\textit{End ~~~~~~~(** amendment algorithm **)}\\
 &$i=i+1$\\
\multicolumn{3}{l}{\textit{End ~~~~~(** iteration i **)}}\\
\multicolumn{3}{l}{sum\_rate, $\check{{\bm t}}$, $\check{{\bm p}}_{S}$ and $\check{{\bm p}}_{R}$ \textit{are the obtained weighted sum rate, feasible subcarrier pairing and power allocation, respectively.}}
\end{tabular}
\end{table}

\begin{table}
\renewcommand{\arraystretch}{1.3}
\caption{Algorithm for the total power constrained case with extra direct-link transmission.}
\label{Table_algo_obtain_feasible_typeII} \centering
\begin{tabular}{lll}
\multicolumn{3}{l}{\textit{Initialize }$i=1, ~\mu^{(1)}, ~{\bm \alpha}^{(1)}, ~\epsilon=0.01$, amendment = \textit{false}, max\_it = $\infty$, sum\_rate = 0}\\
 \multicolumn{3}{l}{\textit{While }($i<$ max\_it) ~~~~~~~\textit{(** iteration i **)}}\\
 &\textit{Compute }$\left(Y^R_{k,m}\right)^{(i)}, \left(Y^D_{k,m}\right)^{(i)}, ~\forall k, m$, \textit{ using }$\mu^{(i)}$ \textit{ in } (\ref{compute Y_R}) \textit { and } (\ref{compute Y_D}).\\
 &$\textit{Obtain }{\bm s}^{(i)}\textit{ using }\left(Y^R_{k,m}\right)^{(i)}, \left(Y^D_{k,m}\right)^{(i)} \textit{ in } (\ref{Eq_s_choose})$\\
 &\textit{Compute }$Y_{k,m}^{(i)}, ~\forall k, m$, \textit{ using }$\left(Y^R_{k,m}\right)^{(i)}$, $\left(Y^D_{k,m}\right)^{(i)}$, ${\bm s}^{(i)}$, ${\bm \alpha}^{(i)}$ \textit{ in } (\ref{compute Y})\\
 &$\textit{Compute }{\bm t}^{(i)}\textit{ using }Y_{k,m}^{(i)} \textit{ in } (\ref{optimal_t_typeII})$\\
 &$\textit{Compute }{\bm p}^{(i)}\textit{ using }\mu^{(i)}, ~{\bm t}^{(i)}, ~{\bm s}^{(i)} \textit{ in }(\ref{optimal power_typeII})$\\
 &$\textit{Compute }\mu^{(i+1)}, ~{\bm \alpha}^{(i+1)}\textit{ using }\mu^{(i)}, ~{\bm \alpha}^{(i)}, ~{\bm t}^{(i)}, ~{\bm p}^{(i)} \textit{ in }(\ref{sub_gradient_typeII_total})$\\
 &\textit{If} (amendment = \textit{false}) \textit{and} $\left(\frac{\lvert\mu^{(i+1)}-\mu^{(i)}\rvert}{\lvert \mu^{(i+1)} \rvert}<\epsilon\right)$ \textit{and} $\left(\frac{\lVert {\bm \alpha}^{(i+1)}- {\bm \alpha}^{(i)}\rVert}{\lVert {\bm \alpha}^{(i+1)} \rVert}<\epsilon\right)$\\
 &\hspace{5 mm}amendment = \textit{true}\\
 &\hspace{5 mm}max\_it = $\lfloor 1.1 \times i \rfloor$\\
 &\textit{End}\\
 &\textit{If} (amendment = \textit{true}) ~~~~~~~\textit{(** amendment algorithm **)}\\
 &\hspace{5 mm}$\hat{{\bm t}} = {\bm t}^{(i)}$, $\hat{{\bm s}} = {\bm s}^{(i)}$, $c_m=\sum_{k=1}^M\hat{t}_{k,m}, \forall m$\\
 &\hspace{5 mm}\textit{For (j = 1 to M)}&\\
 &\hspace{7 mm}\textit{If } $(c_j>1)$, ~~$s^*=\arg\max_{\left\{s|\hat{t}_{s,j}=1\right\}}Y_{s,j}^{(i)}$, ~~\textit{End}\\
 &\hspace{7 mm}\textit{While }($c_j>1$) \\
 &\hspace{10 mm}$m^*=\arg\min_{\{m|c_m=0\}}\left|\alpha^{(i)}_j-\alpha^{(i)}_m\right|$\\
 &\hspace{10 mm}$r^*=\arg\max_{\left\{r|\hat{t}_{r,j}=1, ~r\neq s^*\right\}}Y_{r,m^*}^{(i)}$\\
 &\hspace{10 mm}$\hat{t}_{r^*,j}=0$, ~$\hat{t}_{r^*,m^*}=1$\\
 &\hspace{10 mm}$c_{j}=c_{j}-1, ~c_{m^*}=c_{m^*}+1$\\
 &\hspace{7 mm}\textit{End}\\
 &\hspace{5 mm}\textit{End}\\
 &\hspace{5 mm}\textit{With fixed subcarrier pairing} $\hat{{\bm t}}$ \textit{and} mode selection $\hat{{\bm s}},$ \textit{ apply water-filling on the direct-link, extra direct-link}\\
 &\hspace{5 mm}~~\textit{subcarriers and the relay mode subcarrier pairs with equivalent channel gains in (\ref{Eq_ECG}), to compute}\\
 &\hspace{5 mm}~~\textit{power allocation }$\hat{{\bm p}}$\textit{ and the weighted sum rate} $R$.\\
 &\hspace{5 mm}\textit{If } ($R>$ sum\_rate), ~~sum\_rate $=R$, ~$\check{{\bm t}}=\hat{{\bm t}}$, ~$\check{{\bm s}}=\hat{{\bm s}}$, ~$\check{{\bm p}}=\hat{{\bm p}}$, ~~\textit{End}\\
 &\textit{End ~~~~~~~(** amendment algorithm **)}\\
 &$i=i+1$\\
\multicolumn{3}{l}{\textit{End ~~~~~(** iteration i **)}}\\
\multicolumn{3}{l}{sum\_rate, $\check{{\bm t}}$, $\check{{\bm s}}$ and $\check{{\bm p}}$ \textit{are the obtained weighted sum rate, feasible subcarrier pairing, mode selection and power allocation.}}
\end{tabular}
\end{table}

\end{document}